\documentclass[prb,showpacs,floatfix,superscriptaddress,twocolumn,aps,10pt,longbibliography,groupedaddress]{revtex4-2}
\usepackage{CJK}
\usepackage[english]{babel} 
\usepackage{bm}
\usepackage{ulem}
\usepackage{graphicx}
\usepackage{amsmath}
\usepackage{amssymb}
\usepackage{wasysym}
\usepackage{comment}
\usepackage[dvipsnames]{xcolor}
\usepackage{tikz}
\usepackage{adjustbox}
\usepackage{booktabs}
\usepackage{array}
\usepackage{multirow}
\usepackage{lipsum}   % For generating dummy text
\usepackage{makecell}
\usepackage{enumitem}   

\usepackage[table,dvipsnames]{xcolor}
\usepackage[colorlinks,linktocpage,bookmarks=false]{hyperref}

\newcommand{\hy}[1]{{\color{black} #1}}
\newcommand{\lu}[1]{{\color{black} #1}}

\newcommand\myshade{85}
\definecolor{myrulecolor}{RGB}{150,20,0}% define the color for the rules
\colorlet{mylinkcolor}{violet}
\colorlet{mycitecolor}{YellowOrange}
\colorlet{myurlcolor}{Aquamarine}
\hypersetup{
	linkcolor  = myurlcolor!\myshade!black,
	citecolor  = mycitecolor!\myshade!black,
	urlcolor   =  myrulecolor!\myshade!black,
}

\begin{document}
\begin{CJK*}{UTF8}{gbsn} % Use default fonts from CJK (see below)
%%%%%%%%%%%%%%%%%%%%%%%%%%%%%%%%%%%%%%%%%%%%%%%%%%%%%%%%%%%%%%%%%%%%%%%%%%%%%%%
\title{Hyperbolic Fracton Model, Subsystem Symmetry and Holography III: Extension to Generic Tessellations}

\author{Yosef Shokeeb}
\email[]{yshokeeb@proton.me}
\affiliation{CNRS, Universit\'e de Bordeaux, LOMA, UMR 5798, 33400 Talence, France}

\author{Ludovic D.C. Jaubert}
\email[]{ludovic.jaubert@u-bordeaux.fr}
\affiliation{CNRS, Universit\'e de Bordeaux, LOMA, UMR 5798, 33400 Talence, France}

\author{Han Yan (闫寒)}
\email{hanyan@issp.u-tokyo.ac.jp}
\affiliation{Institute for Solid State Physics, The University of Tokyo, Kashiwa, Chiba 277-8581, Japan}
%%%%%%%%%%%%%%%%%%%%%%%%%%%%%%%%%%%%%%%%%%%%%%%%%%%%%%%%%%%%%%%%%%%%%%%%%%%%%%%
\begin{abstract}
We generalize the Hyperbolic Fracton Model from the $\{5,4\}$ tessellation to generic \lu{\{$p,q$\}} tessellations, and investigate its core properties -- \lu{ground-state degeneracy}, fracton mobility, and holographic correspondence -- \lu{bringing to light a} richer and more intricate structure \hy{than the original case}. 
The ground-state degeneracy and subsystem symmetries are \lu{computed exactly through the inflation rule, but \hy{do not admit} a simple, \hy{symmetric} pattern. For %\hyd{small connectivity} 
$q=3$, we find that the degeneracy is finite, either unique or four-fold. \hy{For the 
\{4,4\} square lattice as the flat limit of $q=4$, 
the degeneracy is well known to be subextensive.}   But for all other tessellations of $q>3$, the degeneracy becomes extensive, including for the \{3,6\} lattice on flat space which maps onto the honeycomb color code.}
The fracton excitation \hy{number follows} exponential-in-distance and algebraic-in-lattice-size growing  patterns when moving outward, and depends sensitively to the tessellation geometry, differing qualitatively from both type-I or type-II fracton model on flat lattices.
Despite this increased complexity, the hallmark holographic features --- subregion duality via Rindler reconstruction, the Ryu-Takayanagi formula for mutual information, and effective black hole entropy scaling with horizon area --- remain valid. 
These results demonstrate that the holographic correspondence in fracton models persists in generic tessellations, and provide a natural platform to explore more intricate subsystem symmetries and  fracton physics.
\end{abstract}
%%%%%%%%%%%%%%%%%%%%%%%%%%%%%%%%%%%%%%%%%%%%%%%%%%%%%%%%%%%%%%%%%%%%%%%%%%%%%%%
\maketitle
\end{CJK*}

\section{Introduction}

\hy{ 
Fracton phases of matter have emerged as an exciting extension of conventional paradigms in condensed matter physics and quantum information theory. Fracton systems host quasiparticle excitations with fundamentally restricted mobility, ranging from immobile fractons to subdimensional particles confined to lines or planes~\cite{ChamonPhysRevLett.94.040402,YoshidaPhysRevB.88.125122,BRAVYI2011839,Haah2011,Vijay2015,Nandkishoreannurev,pretko2020fractonReview},
and motivate  a wide range of research in quantum error correction codes, new phases of matter, and constrained dynamics. 
These unconventional properties reflect an intimate interplay between many-body entanglement, lattice geometry, and generalized symmetries acting on lower-dimensional subsystems~\cite{Shirley2018PhysRevX.8.031051,SLAGLE2019AOP,YanSlagle2022arXiv}.
}

\hy{
A distinctive feature of fracton physics is its  sensitivity to lattice geometry beyond topology alone. Unlike conventional topological phases  whose universal properties are largely insensitive to lattice details, fracton models can exhibit qualitatively different behavior when formulated on lattices with different  coordination, curvature, or dimensional embedding. This geometric dependence has led to fruitful connections with elasticity theory, higher-rank gauge theories, and, in certain limits, structures reminiscent of   gravity~\cite{Pretko2017PhysRevD,Xu2006PhysRevB.74.224433,Yan20a}. As a result, fracton systems provide a natural arena in which concepts from quantum information and condensed matter intersect with ideas traditionally associated with high-energy physics.}

\hy{In parallel, the holographic principle -- most prominently realized in the anti-de Sitter/conformal field theory (AdS/CFT) correspondence~\cite{Maldacena1999,Gubser1998PLB,Witten1998} -- has inspired a broad search for simplified, non-string-theoretic toy models that capture selected aspects of holography~\cite{Susskind1995arXiv,hooft2009arxiv,Zaanen2015_ADSCMTbook,Hartnoll:2018xxg,breuckmann2020critical,Swingle2012,Pastawski2015,Almheiri2015,Yang2016,alex2019majorana,Jahneaaw0092,Freedman2017CMaPh}. While fracton models are not gravitational theories, their constrained dynamics, subregion symmetries, and sensitivity to geometry make them appealing candidates for exploring holography-inspired phenomena in a controlled many-body setting~\cite{Yan2019PhysRevBfracton1,Yan2019PhysRevBfracton2,Yan2020PhysRevBGeodesicString}.
}

\hy{
Motivated by these perspectives, in Refs.~\cite{Yan2019PhysRevBfracton1,Yan2019PhysRevBfracton2,Yan2020PhysRevBGeodesicString} we introduced and studied the  Hyperbolic Fracton Model (HFM), the simplest classical fracton model defined on a regular hyperbolic lattice, specifically the $\{5,4\}$ tessellation. Hyperbolic lattices provide a discrete realization of negatively curved geometry and naturally implement a hierarchical bulk-boundary structure~\cite{BoylePhysRevX.10.011009}. Despite its simplicity, the HFM exhibits a number of remarkable properties, including extensive ground-state degeneracy, emergent boundary degrees of freedom, and geodesic-like constraints that encode bulk-boundary relations. Subsequent work further clarified these features and established connections to $p$-adic and tensor-network-inspired toy models of holography~\cite{Yan_Jepsen_Oz_2025,duran2024conformalboundaryholographicdual}.
}

Here we generalize the original \lu{classical} HFM to all $\{p,q\}$ tessellations of the hyperbolic plane, extending beyond the original simple case, and discover \lu{unconventional} behavior without analogue in the type-I and type-II fracton models in flat 3D lattices~\cite{PhysRevB.94.235157}. The construction of our model is based on a set of geometry-dependent, recursive inflation rules that dictate the subsystem symmetries of the system. Through a detailed analysis of the model's ground state degeneracy and properties, we derive scaling laws governed by the tessellations' parameters $p$ and $q$. \lu{We find that almost all tessellations with $q>3$ support an extensive ground-state degeneracy. In particular the HFM on the Euclidean \{3,6\} lattice is equivalent to the classical limit of the honeycomb color code with a $\mathbb{Z}_2$ gapped topological order phase \cite{PhysRevLett.97.180501}.} Another central result of our generalized model is the Rindler reconstruction, a procedure through which bulk spin configurations within a minimal wedge are uniquely determined by measurements on a corresponding boundary subregion. This provides a discrete analogue to the entanglement wedge reconstruction in continuum AdS/CFT. We also demonstrate that the mutual information between complementary boundary subregions, which is the classical analogue of entanglement entropy, follows a discrete form of the Ryu-Takayanagi formula~\cite{Ryu2006PhysRevLett}, therefore linking boundary entanglement to the length of the shared boundary. Furthermore, we establish that the removal of bulk spins simulates the formation of a black hole. This induces a change in entropy proportional to the boundary area of the defect region, and offers a discrete analogue of the Bekenstein-Hawking entropy formula~\cite{Hawking_1975}.
Finally, we have made quantitative analysis of how fractons \lu{grow} exponentially in distance and algebraically in system size when pushed away to \lu{outwards} layers. 
This feature is qualitatively different from \lu{both type-I and} type-II fracton models on flat cubic lattice, and applies to quantum fracton orders defined on lattices embedded in 2D hyperbolic space $\times$ 1D flat space too \cite{yan2022ycube}.

This paper is structured as follows: Section~\ref{sec:ads-cft} provides a brief review of the essential concepts of the AdS/CFT correspondence that underpin our model. Section~\ref{sec:lattice} elaborates on the recursive construction of generic hyperbolic tessellation geometry. Section~\ref{sec:fracton-model} introduces the Plaquette Ising model as a foundation for the hyperbolic generalization and presents the Hyperbolic Fracton Model~\cite{Yan2019PhysRevBfracton1,Yan2019PhysRevBfracton2}. The following sections present our results. In Section~\ref{sec:gsd}, we derive \lu{exact} expressions for the ground state degeneracy and residual entropy. Then, the holographic properties of the model are elaborated through the Rindler reconstruction in Section~\ref{sec:rindler_reconstruction}, mutual information calculations in Section~\ref{sec:mutual}, and an analysis of black hole entropy in Section~\ref{sec:BH}. Stepping away from the ground-state properties, section~\ref{sec:fractons} is devoted to the analysis of fracton excitations.
Finally we conclude our paper with a discussion in Section~\ref{sec:conc}.
 
\section{AdS/CFT correspondence}
\label{sec:ads-cft}

The holographic principle states that the information in quantum gravity that describes a volume of space can be fully encoded on its boundary region. It originated in the study of black hole thermodynamics, which attempts to reconcile general relativity with quantum mechanics through the study of black holes. The resolution of the black hole information paradox led to the formulation of the Bekenstein-Hawking entropy formula~\cite{Hawking_1975}:
$$ S_{\text{BH}} = \frac{A}{4G} $$
where $A$ is the area of the black hole horizon and $G$ is the gravitational constant. This result implies that the entropy of a black hole scales not with its volume but with its surface area, which directly inspired the holographic principle.

The first concrete realization of the holographic principle is the AdS/CFT correspondence~\cite{Maldacena1999}, which is a conjectured duality between a $(d+1)$-dimensional gravitational theory in an Anti-de Sitter (AdS) spacetime and a $d$-dimensional conformal field theory (CFT) on its boundary. AdS space is a maximally symmetric Lorentzian manifold with negative curvature. %\hyd{For example, $H^2\times\mathbb{R}$ is the universal cover of AdS$_3$, meaning that AdS$_3$ is locally equivalent to a temporal stack of two-dimensional hyperbolic planes $H^2$.}
Meanwhile, a CFT is a field theory that is invariant under conformal symmetries, particularly powerful for describing systems at their critical points. The AdS/CFT duality is a \textit{strong-weak duality}, where the boundary CFT that is strongly coupled and analytically intractable, corresponds to a weakly coupled gravitational theory in the bulk.

A cornerstone of AdS/CFT is the Ryu-Takayanagi (RT) formula, which relates the entanglement entropy in the boundary CFT to the geometry of the bulk AdS space. For a spatial region $A$ in the CFT, the entanglement entropy $S_A$ is proportional to the area of the minimal  bulk surface $\gamma_A$  homologous to $A$ (i.e., the minimal surface's boundary is also boundary of $A$):
$$ S_A = \frac{\text{Area}(\gamma_A)}{4G} $$
This formula provides a geometric interpretation of quantum entanglement and generalizes the Bekenstein-Hawking formula. As we will demonstrate in Section~\ref{sec:mutual}, the HFM reproduces a discrete version of this formula, where mutual information serves as a classical analogue for the entanglement entropy.

\hy{In our work, we will focus on a constant time slice of AdS$_3$, which has the geometry of a hyperbolic plane $\mathbb{H}^2$ with constant, negative curvature. 
}
\section{Hyperbolic Tessellations}
\label{sec:lattice}

%==================
\begin{figure}
    \centering
    \includegraphics[width=1\linewidth]{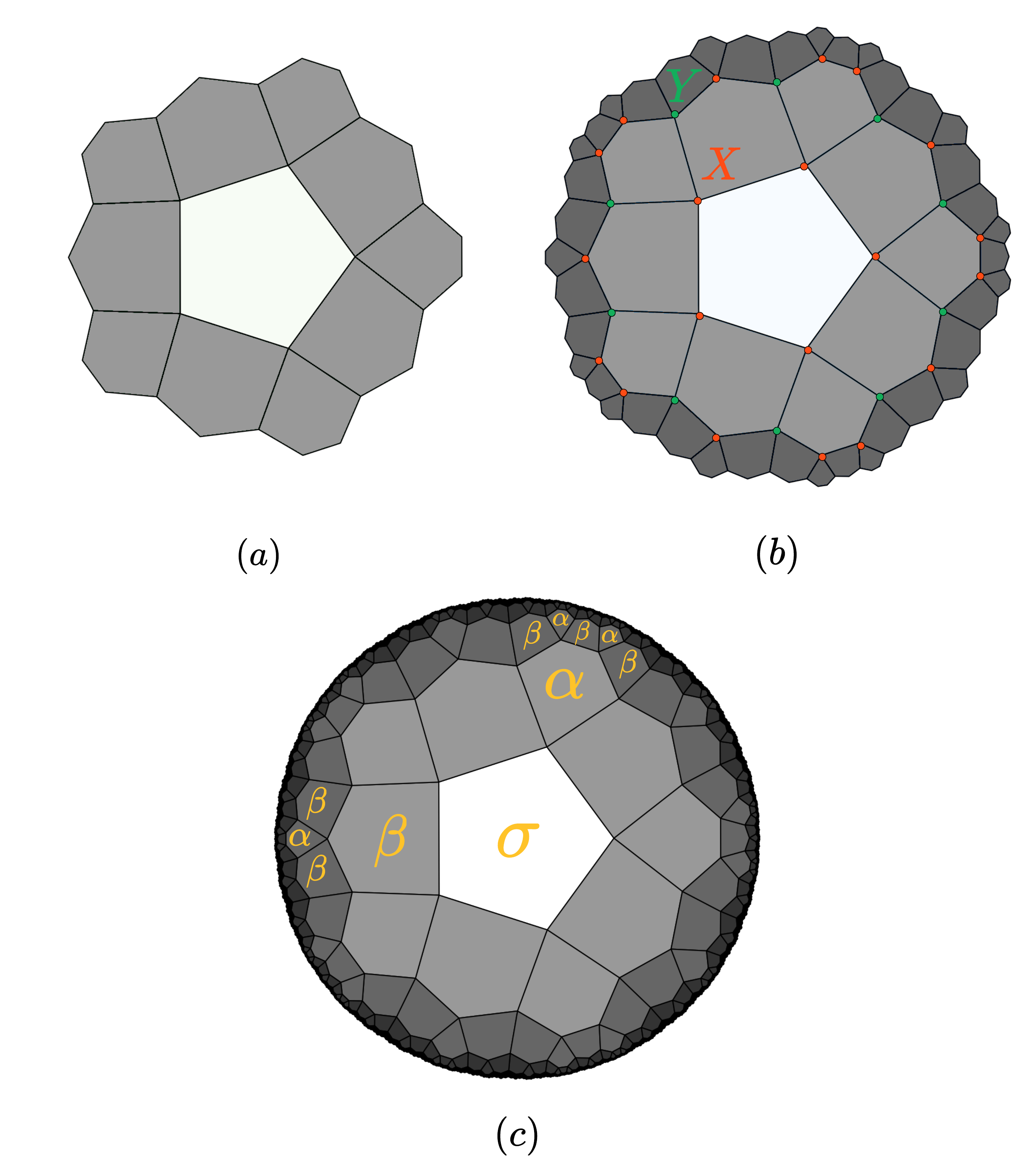}
    \caption{
    Layer-by-layer construction of a $\{5,4\}$ hyperbolic tessellation. Each new layer's generation  follows an inflation rule. $(a)$ one layer, $(b)$ two layers, $(c)$ three layers. The different types of polygons ($\alpha, \beta, \sigma$) and vertices ($X, Y$) are labeled according to their connectivity to the previous layer. The recursive inflation rules generate a self-similar structure where the number of polygons grows exponentially with the number of layers $l$.
    }
    \label{fig:lattice}
\end{figure}
%==================

A regular tessellation is a tiling of a surface by identical, regular polygonal faces, where the same number of faces meet at each vertex. These structures are uniquely classified by their Schl\"afli symbol $\{p,q\}$, where $p$ is the number of edges of each polygonal face, and $q$ is the number of faces meeting at each vertex of the tiling. The geometry of the embedding space is determined by the relationship between $p$ and $q$. For a flat Euclidean plane, tessellations must satisfy $1/p + 1/q = 1/2$, which permits only three configurations: the square tiling $\{4,4\}$, the triangular tiling $\{3,6\}$, and the hexagonal tiling $\{6,3\}$. In contrast, tilings of the hyperbolic plane \lu{negatively curved space} corresponds to the inequality $1/p + 1/q < 1/2$, which allows for an infinite number of regular tessellations.

\hy{All $\{p,q\}$ tessellations shown in this work are drawn using the conformal Poincar\'e disk model of the hyperbolic plane $\mathbb{H}^2$, consistent with the representations used in earlier papers. In this model, hyperbolic geodesics are represented by circular arcs orthogonal to the boundary of the disk. The apparent straight edges in some figures do not correspond to a change of hyperbolic representation (e.g., to the Klein model) -- it is merely for better visualization.}

In the main body of this work, we shall focus on the class of tessellations with $q > 3$.
\lu{As we will  discuss in section \ref{app:q=3}, models} with $q=3$ are strongly constrained \lu{by the trigonal symmetry of the vertices} and have a different behavior from other tessellation. 

The construction of our lattices \lu{for $q>3$} follows a set of \textit{inflation rules}~\cite{BoylePhysRevX.10.011009}, growing the lattice layer-by-layer ($k=1, 2, \ldots$) from a central polygon, denoted $\sigma$, as depicted in Figure~\ref{fig:lattice}. To formalize the construction, we classify the polygons and vertices at each layer based on their connectivity to the preceding layer:
\begin{itemize}
    \item $\alpha$-polygons: They share a single vertex with polygons of the previous layer.
    \item $\beta$-polygons: They share two vertices with polygons of the previous layer.
    \item $X$-vertices: They connect polygons in the new layer to one polygon in the previous layer.
    \item $Y$-vertices: They connect polygons in the new layer to two polygons in the previous layer.
\end{itemize}
The inflation rules $\tau$ specify how $\alpha$- and $\beta$-polygons at a given layer generate a sequence of descendant polygons in the next layer, therefore ensuring that the symmetries of the tessellations are preserved at each step~\cite{BoylePhysRevX.10.011009}:
\begin{align}
    \tau: \quad \alpha &\rightarrow \alpha^{r} (\beta \alpha^{q-3})^{p-3} \beta \alpha^{r} \nonumber\\
    \beta &\rightarrow \alpha^{r} (\beta \alpha^{q-3})^{p-4} \beta \alpha^{r}
    \label{eq:inflationab}
\end{align}
where $r = (q-4)/2$. The inflation process can be described by an inflation matrix $\mathbf{M}_{\tau}$ discussed in detail in the following, whose eigenvalues dictate the exponential growth rate of the number of polygons and vertices.

\subsection{Polygon Inflation}
\label{app:inflation}

This subsection provides a detailed framework for the recursive construction of the hyperbolic tessellations used in this work. The growth of the lattice is governed by a set of linear recurrence relations that can be expressed using an \textit{inflation matrix} $\mathbf{M}_{\tau}$. The diagonalization of this matrix is crucial for deriving the analytical expressions for the ground state degeneracy, residual entropy, and black hole entropy presented in this work.

For a generic $\{p,q\}$ tessellation with $q>3$, the polygons at each layer can be classified into $\alpha$-polygons and $\beta$-polygons based on their connectivity to the previous layer. We can represent the number of each type of polygon at layer $k$ as a vector, ${\mathcal{N}^{k}} = (N^k_\alpha, N^k_\beta)^T$. The layer-to-layer growth is governed by the polygon inflation matrix $\mathbf{M}_{\tau}$:
\begin{eqnarray}
\begin{split}
&\mathbf{M}_{\tau} =\\
&
\begin{pmatrix}
     (q-4) + (q-3)(p-3) & (q-4) + (q-3)(p-4) \\ 
     (p-2) & (p-3)
\end{pmatrix}
\end{split}
\end{eqnarray}
Starting from the central polygon ($\sigma$-polygon), the configuration at the first layer is ${\mathcal{N}^{1}} = (p(q-3), p)^T$. Subsequent layers are generated by the recurrence relation:
\begin{eqnarray}
    {\mathcal{N}^{k+1}} = \mathbf{M}_\tau {\mathcal{N}^{k}}
\end{eqnarray}
The total number of polygons, $N_p$, on a lattice with $l$ layers is the sum of polygons at each layer, plus the initial $\sigma$-polygon:
\begin{eqnarray}
    N_p = 1 + \sum_{k=1}^{l} (N^k_{\alpha} + N^k_{\beta})
\end{eqnarray}
\subsection{Vertex Inflation}
Similarly, the growth of vertices can be described by an inflation process. For $p,q > 3$, the vertices are classified as X-vertices or Y-vertices. Their recursive growth is governed by the same inflation rules, $\tau$, that dictate the polygon evolution:
\begin{align*}
\tau: \quad &X \to X^{\frac{q-4}{2}}\left(YX^{q-3}\right)^{p-3}YX^{\frac{q-4}{2}} \\
&Y \to X^{\frac{q-4}{2}}\left(YX^{q-3}\right)^{p-4}YX^{\frac{q-4}{2}}
\end{align*}
The number of vertices of each type at layer $k$, denoted by the vector ${\mathcal{V}^k} = (N^k_X, N^k_Y)^T$, also follows the same matrix evolution, ${\mathcal{V}^{k+1}} = \mathbf{M}_\tau{\mathcal{V}^k}$. Starting with the vertices of the central polygon, the initial vector is ${\mathcal{V}^1} = (p, 0)^T$. The total number of vertices, $N_v$, on a lattice with $l$ layers is:
\begin{eqnarray}
    N_v = \sum_{k=1}^{l} (N^k_X + N^k_Y)
\end{eqnarray}
\subsection{The Inflation Matrix}
\label{sec:inflationmatrix}
To find a closed-form expression for $N_p$ and $N_v$, we need to compute powers of the inflation matrix, $\mathbf{M}_{\tau}^k$. This is done by diagonalizing the matrix:
\begin{eqnarray}
\mathbf{M}_{\tau} = \mathbf{P} \mathbf{D}_\tau \mathbf{P}^{-1}
\end{eqnarray}
where $\mathbf{D}_\tau$ is the diagonal matrix of eigenvalues and $\mathbf{P}$ is the matrix of corresponding eigenvectors. The eigenvalues are:
\begin{eqnarray}
\lambda_{\pm} = \frac{p(q-2)}{2} - (q-1) \pm \delta
\end{eqnarray}
where
\begin{eqnarray}
     \delta = \frac{1}{2}\sqrt{(p(q-2) -2q +2)^2 -4}
\end{eqnarray}
The eigenvector and inverse eigenvector matrices are:
\begin{align}
\mathbf{P} &= 
\begin{pmatrix}
\cfrac{(p-2)(q-4) -2\delta}{2(p-2)} & \cfrac{(p-2)(q-4) + 2\delta}{2(p-2)} \\
1 & 1
\end{pmatrix} \\
\mathbf{P}^{-1} &= 
 \begin{pmatrix}
     -\cfrac{p-2}{2\delta} & \cfrac{(p-2)(q-4) + 2\delta}{4\delta} \\ 
     \cfrac{p-2}{2\delta} & -\cfrac{(p-2)(q-4) - 2\delta}{4\delta}
\end{pmatrix} 
\end{align}
The $k$-th power of the matrix is then given by $\mathbf{M}_{\tau}^k = \mathbf{P} \mathbf{D}_\tau^k \mathbf{P}^{-1}$. Setting $P = p-2$ and $Q = q-4$, one gets
\begin{widetext}
\begin{eqnarray}
    \mathbf{M}_\tau^k = 
    \frac{1}{4\delta}\begin{pmatrix}
    (PQ + 2\delta)\lambda_+^k - (PQ - 2\delta)\lambda_-^k 
    & \frac{(PQ)^2 - 4\delta^2}{2P}(\lambda_-^k - \lambda_+^k) 
    \\
     2P(\lambda_+^k - \lambda_-^k) 
     & (PQ + 2\delta)\lambda_-^k - (PQ - 2\delta)\lambda_+^k
\end{pmatrix},
\end{eqnarray}

\end{widetext}

which can be used to evaluate $\mathcal{N}^{k} = (N^{k}_\alpha, N^{k}_\beta)^T = \mathbf{M}_\tau^{k-1}\, \mathcal{N}^{1}$:
\begin{eqnarray}
    N^k_{\alpha} = A_+ \lambda_+^{k-1} + A_-\lambda_-^{k-1} \label{eq:naal}\\
    N^k_{\beta}  = B_+ \lambda_+^{k-1} + B_-\lambda_-^{k-1} \label{eq:nbbl}
\end{eqnarray}
with
\begin{align}
    A_+ &= \frac{p ((p-2)(q-4)+2\delta)((p-2)(q-2)+2\delta)}{8\delta(p-2)}, \\
    A_- &= -\frac{p ((p-2)(q-4)-2\delta)((p-2)(q-2)-2\delta)}{8\delta(p-2)}, \\
    B_+ &= \frac{p}{4\delta}\left[2\delta+(p-2)(q-2)\right], \\
    B_- &= \frac{p}{4\delta}\left[2\delta - (p-2)(q-2)\right].
\end{align}
In the thermodynamic limit, we obtain the ratio $R_\infty$ that will be used in Sec. \ref{sec:gsd}
\begin{eqnarray}
    R_\infty = \lim_{k\to\infty} \frac{N^{k}_\beta}{N^{k}_\alpha} = \frac{B_+}{A_+}=\frac{2P}{PQ+2\delta}.
    \label{eq:Rinf}
\end{eqnarray}

\section{The Fracton Model: From Euclidean to Hyperbolic Lattices}
\label{sec:fracton-model}

%------------------------------------------------------------
\subsection{Plaquette Ising Model}\label{sec:efm}
To establish the framework for the hyperbolic fracton model, we first review the foundational concepts of fracton physics through the classical toy model: the Plaquette Ising Model (PIM) \lu{on the \{4,4\} tessellation}.

%==================
\begin{figure}
    \centering
    \includegraphics[width=1\linewidth]{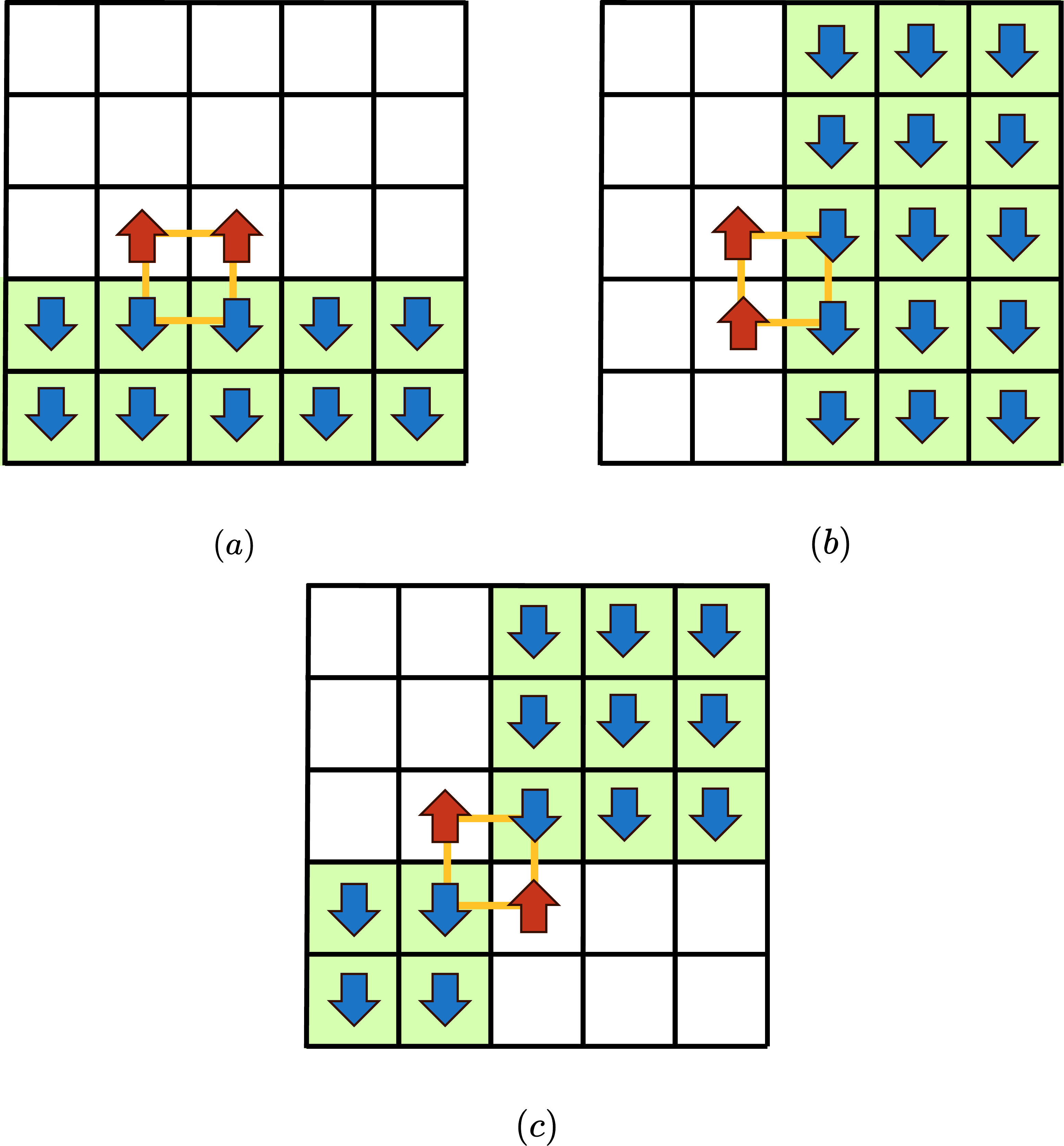}
    \caption{The subsystem symmetries of the Plaquette Ising Model. Starting from an all-spins-up ground state and flipping spins along $(a)$ horizontal, $(b)$ vertical lines, or a combination of both $(c)$ will result in another ground state. These operations form the basis of the model's sub-extensive ground state degeneracy.}
    \label{fig:plaquette_gs}
\end{figure}
%==================

In the PIM, classical Ising spins $S_i^z = \pm 1$ reside at the centers of each square, while the interactions are defined at the vertices. For each vertex $v$ shared by four adjacent squares, we define a vertex operator:
\begin{gather*}
    \mathcal{O}_v = \prod_{i=1}^{q} S_{i}^{z}
\end{gather*}
with $q=4$ here. The Hamiltonian of the system is the sum over all such vertex terms:
\begin{gather}\label{eq:hamilton}
    \mathcal{H} = -\sum_{v} \mathcal{O}_v
\end{gather}
The ground states of the model satisfy $\mathcal{O}_v = 1$ for all vertices, with a degeneracy that scales as $\Omega = 2^{L_x + L_y - 1}$ on a torus. These symmetries correspond to flipping all spins along any  row or column (Fig.~\ref{fig:plaquette_gs}), therefore preserving all constraints $\mathcal{O}_v = 1$ while generating new ground states. The entropy $S = k_B \log \Omega \propto k_B \log 2 \times (L_x + L_y)$ scales with the boundary size, which is a signature of fracton order.

\begin{figure}
    \centering
    \includegraphics[width=1\linewidth]{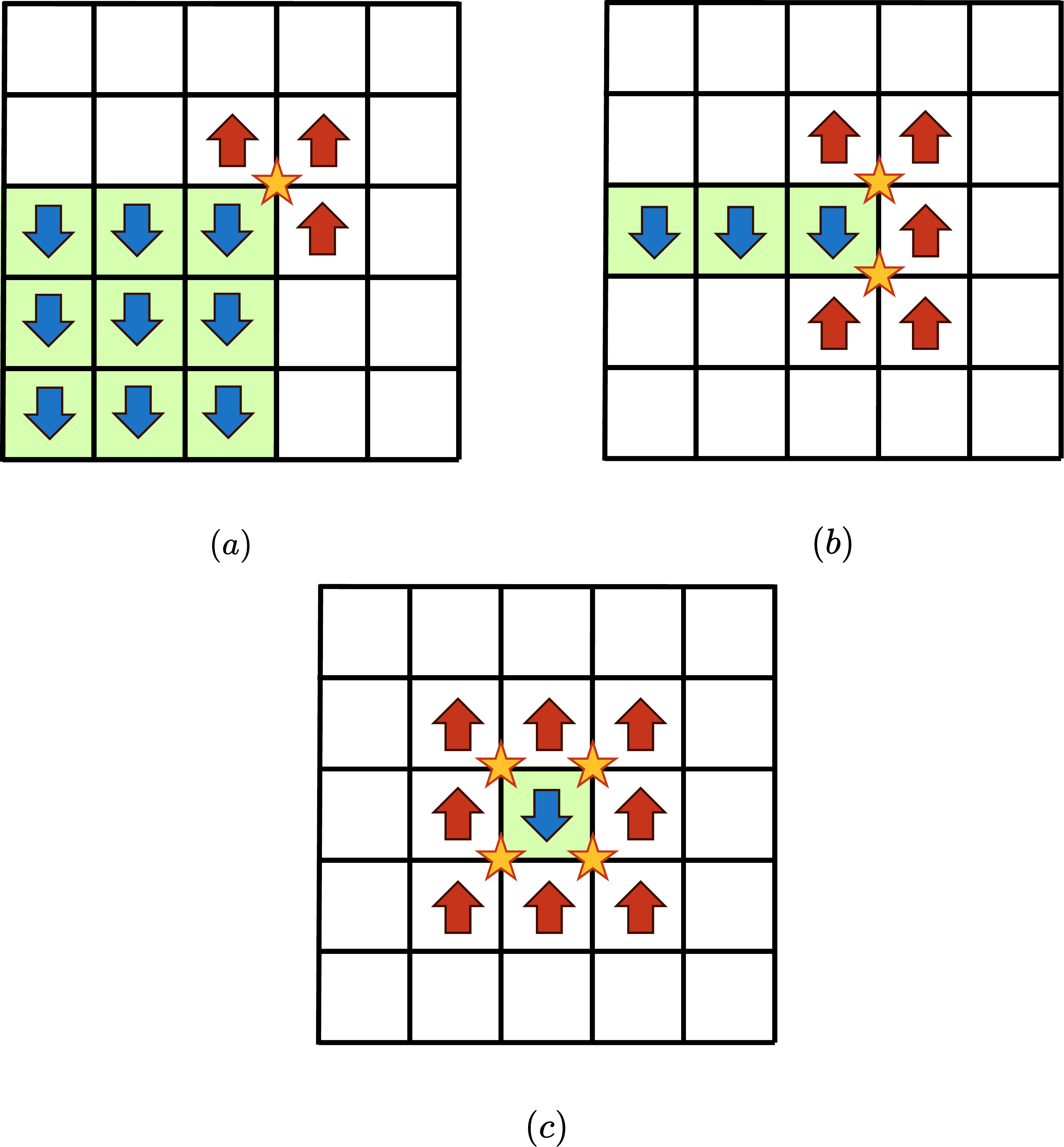}
    \caption{Fracton excitations in the Plaquette Ising Model. $(a)$ A single fracton excitation (represented by a star) is immobile due to the need to flip an infinite number of spins. $(b)$ A line of spin flips creates a bound pair of fractons that can propagate collectively along that line. $(c)$ A single spin flip creates four bound fractons, which can propagate freely.}
    \label{fig:plaquette_fractons}
\end{figure}

Here, the classical fracton excitations correspond to violated constraints $\mathcal{O}_v = -1$. A single fracton is an immobile topological excitation, since any local attempt to move it necessarily creates additional excitations (Fig.~\ref{fig:plaquette_fractons}a). Composite fracton excitations have more mobility. A bound fracton pair (Fig.~\ref{fig:plaquette_fractons}b) can propagate along a line perpendicular to the pair via local spin flips, while a four-fracton quadrupole (Fig.~\ref{fig:plaquette_fractons}c) can move freely on the lattice.

%==================
%==================

%------------------------------------------------------------
\subsection{Hyperbolic Fracton Model}
\label{sec:hfm}

\begin{figure}
    \centering
    \includegraphics[width=1\linewidth]{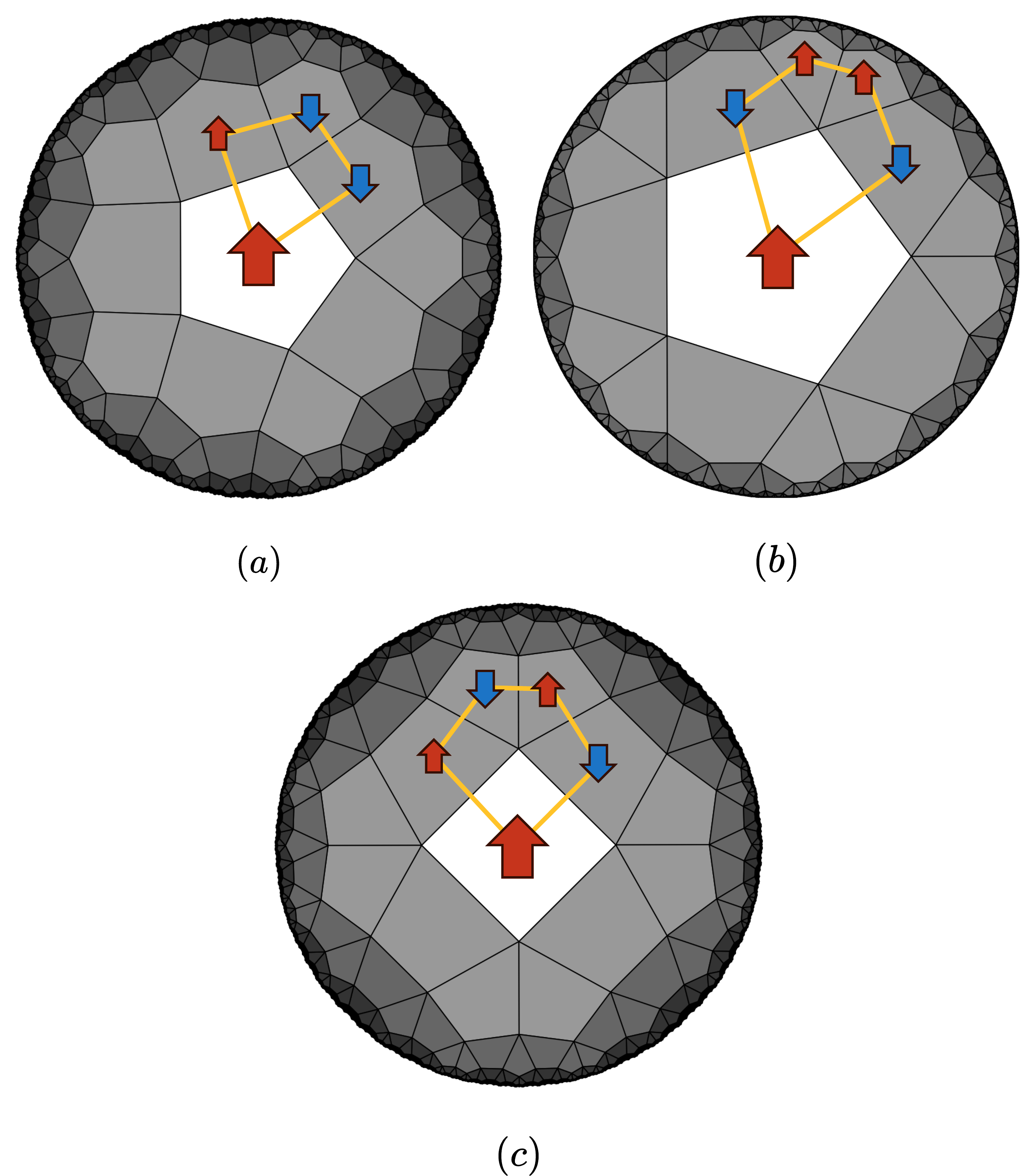}
    \caption{Vertex operators $\mathcal{O}_v$ in the Hyperbolic Fracton Model for various tessellations. The operator is a product of the spins on the polygons meeting at that vertex. $(a)$ A $\{5,4\}$ tessellation has 4-spin operators. $(b)$ A $\{5,5\}$ tessellation has 5-spin operators. $(c)$ A $\{4,5\}$ tessellation also has 5-spin operators.}
    \label{fig:model}
\end{figure}

The Hyperbolic Fracton Model, originally introduced for the $\{5,4\}$ tessellation~\cite{Yan2019PhysRevBfracton1,Yan2019PhysRevBfracton2}, generalizes the plaquette Ising model to generic $\{p,q\}$ hyperbolic tessellations. As in the PIM, \lu{classical} Ising spins $S_i^z$ reside at the center of each $p$-gon and $q$ polygons meet at vertices. The Hamiltonian~\eqref{eq:hamilton} retains its form but vertex operators extend to \lu{$q$} spins (Fig.~\ref{fig:model}).

In the original HFM model~\cite{Yan2019PhysRevBfracton1,Yan2019PhysRevBfracton2}, geodesics and their associated subsystem symmetries play a fundamental role in establishing holographic properties. On the Euclidean lattice, geodesics correspond to straight lines along the square lattice edges. Meanwhile, in the hyperbolic $\{5,4\}$ lattice geodesics manifest as circular arcs that orthogonally intersect the boundary (Fig.~\ref{fig:model}a). These subsystem symmetries were used to demonstrate that the HFM exhibits several key properties of holography \cite{Yan2019PhysRevBfracton1,duran2024conformalboundaryholographicdual}. First, the model realizes the Rindler reconstruction, which states that spin configurations on any connected boundary subregion determine bulk states within their minimal geodesic wedge (the discrete analog of an entanglement wedge). Second, the mutual information for bipartitions of connected boundary subregions satisfies the Ryu-Takayanagi formula. Finally, a naively defined black hole in the model has an entropy that follows the Bekenstein-Hawking formula. 

In hyperbolic lattices with $q > 4$, the subsystem symmetries become more \lu{complex}. 
For odd $q$, there are no geodesics formed by the edges of the lattice, and for even $q$, the geodesics do not define subsystem symmetry directly \cite{Yan_Jepsen_Oz_2025}. 
%geodesics become ill-defined due to the increasing vertex connectivity which limits the analysis of the model through subsystem symmetries. 
The lack of systematic construction of subsystem symmetries motivate  our generalization of the HFM through the use of inflation rules (Section~\ref{sec:lattice}), as they provide an alternative to construct subsystem symmetries and demonstrate holographic properties. Crucially, in section~\ref{sec:gsd} we demonstrate that  the original HFM exhibits an \textit{extensive} entropy that scales with the system's volume. This stands in contrast with the \lu{Plaquette Ising Model}, where subsystem symmetries yield a sub-extensive degeneracy.

%==================
%==================

\section{Ground State Degeneracy}\label{sec:gsd}

%==========================================================
\subsection{General Tessellations}
Let us now compute the \lu{classical} ground state degeneracy (GSD) of a general HFM.
We shall respectively define $N_\alpha^k$, $N_\beta^k$ and $N_p^k$, the number of $\alpha$ and $\beta$ polygons and the total number of polygons (\textit{i.e.}, spins) on layer $k$. $N_v^k$ is the number of vertices between layers $k$ and $k-1$. By definition for $k\geq 1$,

\begin{eqnarray}
    N_p^k=N_\alpha^k + N_\beta^k \qquad \mathrm{and} \qquad N_\beta^k=N_v^k.
    \label{eq:Npabv}
\end{eqnarray}

\begin{figure}
    \centering
    \includegraphics[width=1\linewidth]{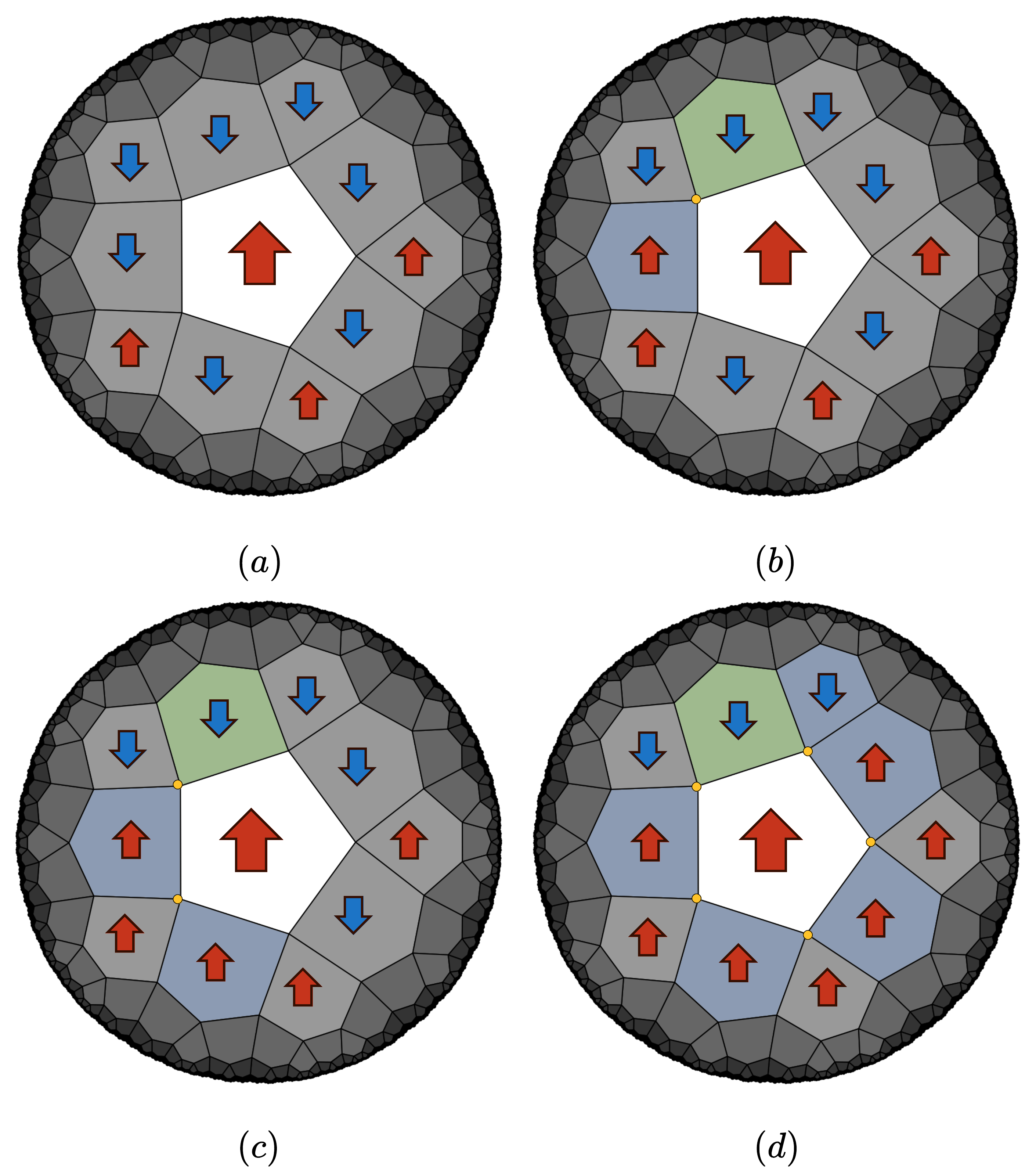}
    \caption{Illustration of the counting procedure on a $\{5,4\}$ tessellation. $(a)$ An initial random configuration. $(b)$ A single spin's value (green polygon) is chosen, introducing one DOF. Its value, combined with the $\mathcal{O}_v=1$ constraints (yellow dots) at adjacent vertices, $(c)$ iteratively fixes the values of neighboring spins (blue polygons). $(d)$ Upon layer closure, a periodic boundary condition removes one DOF. Subsequent layers inherit constraints from prior ones via inflation rules, ensuring consistency across the lattice. The total number of DOF is given by counting the spins left to fluctuate.}
    \label{fig:counting_dof}
\end{figure}

The ground state of the HFM arises from the number of spins free to fluctuate collectively after all constraints on the lattice are satisfied. To calculate the GSD, we shall take advantage of the local $\mathbb{Z}_2$ symmetry of our model. Around a given vertex, the local rule $\mathcal{O}_v=+1$ is necessarily respected if we arbitrarily choose the orientation of $q-1$ spins, whose product will give $\eta=\pm 1$, and then impose the value $\eta$ on the last, $q^{\rm th}$ spin, so that $\mathcal{O}_v=\eta^2=+1$. Let us apply this method recursively (Fig.~\ref{fig:counting_dof}), and compute the contribution of each layer $k$ to the GSD. \\

%%%%%%%%%%%%%%%%%%%
%%%%%%%%%%%%%%%%%%%

%%%%%%%%%%%%%%%%%%%
\begin{figure*}
    \centering
    \includegraphics[width=1\linewidth]{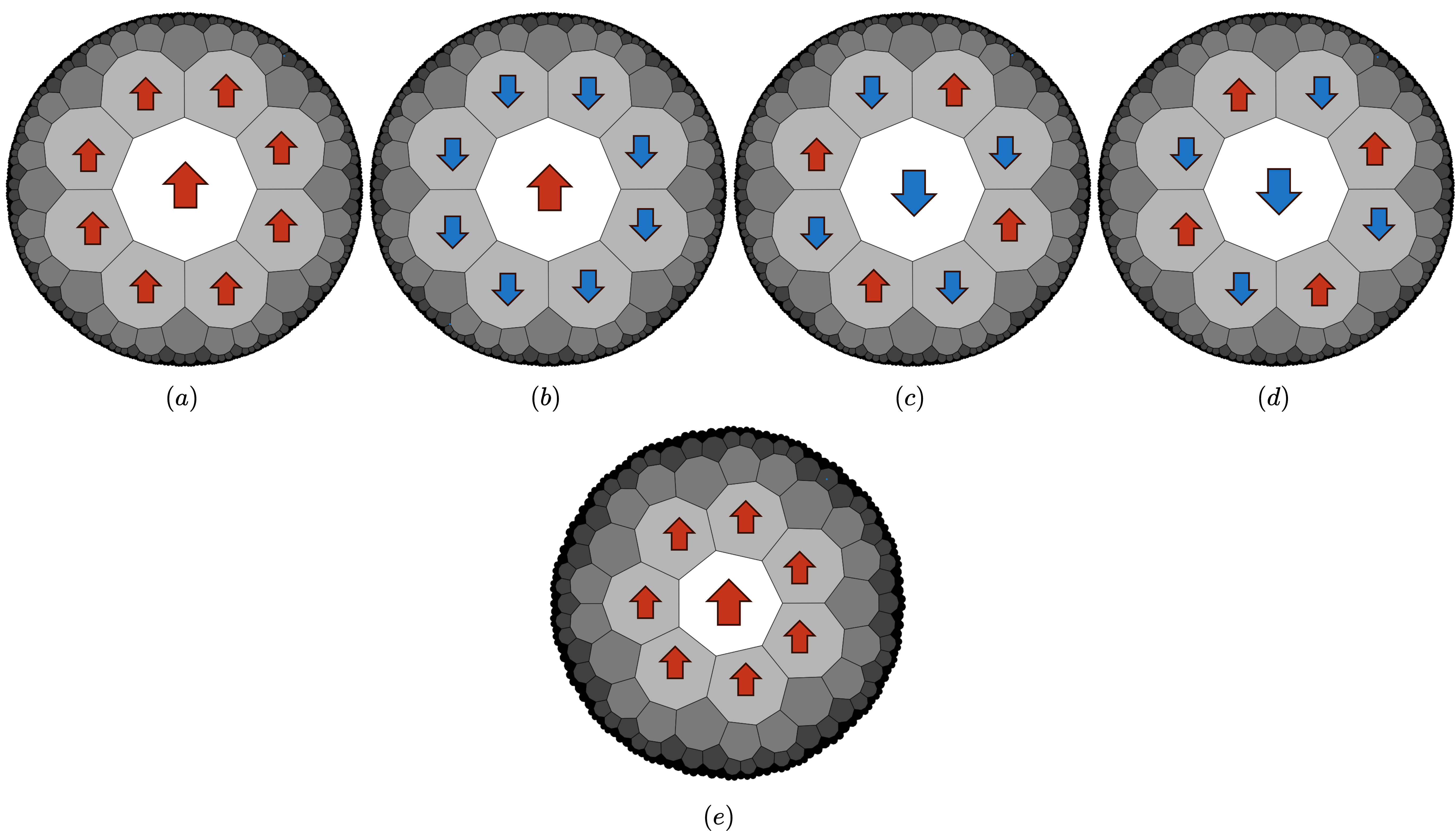}[h]
    \caption{Illustration of ground states for $\{p,3\}$ tessellations. $(a-d)$ The four degenerate ground states of the $\{8,3\}$ tessellation (even $p$). The degeneracy arises from a combination of the central spin's orientation and the configuration of the first layer. $(e)$ The unique ground state of the $\{7,3\}$ tessellation (odd $p$), where the lattice geometry removes all degrees of freedom.}
    \label{fig:q3-tess}
\end{figure*}
%%%%%%%%%%%%%%%%%%%

On the central spin $k=0$, we have two possibilities, $\pm 1$. The degeneracy of layer $k=0$ is thus 2. Let us arbitrarily choose one orientation, and assign an arbitrary value to a spin on a $\beta$ polygon on layer $k=1$ (say the green polygon in Fig.~\ref{fig:counting_dof}.(b)) and consider the vertices and polygons counter-clockwise. We start with a vertex where the orientations of two spins are already known, and there are $(q-2)$ spins left to consider. We can arbitrarily choose the orientation of $(q-3)$ remaining spins, and still respect $\mathcal{O}_v=+1$ thanks to the $q^{\rm th}$ spin. By definition, the $(q-3)$ remaining spins are all $\alpha$ spins, while the final spin that has no freedom is a $\beta$ spin (the blue polygon in Fig.~\ref{fig:counting_dof}.(b)). Repeating the operation one more time, we have again $(q-3)$ $\alpha$ spins with an arbitrary orientation, and one final $\beta$ spin whose orientation is fixed (the second blue polygon in Fig.~\ref{fig:counting_dof}.(c)). Closing the layer $k=1$ counter-clockwise, we repeat the procedure until hitting the initial $\beta$ spin. In this case, since this $\beta$ spin has already been fixed at the beginning, we have to fix one of the $\alpha$ spin in the last procedure in order to respect $\mathcal{O}_v=+1$. As a result, the total number of spins free to fluctuate in the ground state in layer $k=1$ contains all $\alpha$ spins except the last one, and the initial $\beta$ spin, i.e. $N_\alpha^1$ spins; while the number of fixed spins is automatically $N_\beta^1$. The degeneracy of layer $k=1$ is $2^{N_\alpha^{1}}$. It is straightforward to apply the same procedure on each layer $k$, and we always find a degeneracy of $2^{N_\alpha^{k}}$. Including the central spin, the GSD of our model with open boundaries is
\begin{eqnarray}
    \Omega = 2^{\left(1 + \sum_{k=1}^{l} N^k_{\alpha}\right)}
    \label{eq:gsd}
\end{eqnarray}
Another way to derive the above equation is the vertex representation where we subtract the number of constraints per layers. Since the number of degrees of freedom (DOF) in a layer is the total number of spins, $N_p^k$, and the number of constraints imposed by each vertex is $N_v^k$, the number of free DOF per layer is $N_p^k-N_v^k=N^k_{\alpha}$ for $k\geq 1$ (see Eqs.~(\ref{eq:Npabv})).  The ground-state entropy is 
\begin{eqnarray}
    S = k_B \log{2} \times \left(\sum_{k=1}^{l} N^k_{\alpha} + 1 \right).
    \label{eq:SNa}
\end{eqnarray}
Once normalized by the total number of spins $N_p = \sum_{k=1}^{l} (N^k_{\alpha} + N^k_{\beta})+1$ and taking the thermodynamic limit, we get the residual ground-state entropy:
\begin{eqnarray}
    s = \lim_{l\to\infty} \frac{S}{N_p} = \frac{k_B\log{2}}{1 + R_\infty},
    \label{eq:sinfty}    
\end{eqnarray}
where we introduce the ratio $R_l = \frac{\sum_{k=1}^{l} N^k_{\beta}}{\sum_{k=1}^{l} N^k_{\alpha}}$ that captures the asymptotic structure of the lattice. In the thermodynamic limit this ratio converges to a nonzero constant determined by the tessellation's Schl\"afli symbol $\{p,q\}$ (see Eq.~(\ref{eq:Rinf})):
\begin{eqnarray}
    R_{\infty} = \frac{2(p-2)}{\sqrt{\left(p(q-2) -2q +2\right)^2 -4} + (p-2)(q-4)}
    \label{eq:Rinfty}
\end{eqnarray}
\lu{This result is valid for all tessellations \{$p,q>3$\}. $R_{\infty}$ is always finite, except for $p=q=4$ where it diverges. This exception corresponds to the well-known sub-extensive entropy of the Plaquette Ising Model  on the Euclidean square lattice. The HFM on negatively curved space with $q>3$ thus always supports an extensive degeneracy, in sharp contrast with the PIM.}

%==========================================================
\subsection{The $\{p,3\}$ Tessellations}
\label{app:q=3}

The case where $q=3$ represents a constrained limit of the HFM. In this geometry, each vertex is shared by exactly three polygons, which alters the recursive structure of the lattice and the nature of its ground state degeneracy. In particular, the inflation rules are distinct due to the different local connectivity. The polygon types are now denoted $\sigma$ (central), $\beta$ (sharing an edge with the previous layer), and $\gamma$ (sharing \lu{three vertices with the previous layer}). The inflation rules, $\tau$, are given by:
\begin{align}
\tau: \quad \sigma &\to \beta^p \nonumber\\
\beta &\to \gamma^{1/2}\beta^{p-5}\gamma^{1/2} \nonumber\\   
\gamma &\to \gamma^{1/2}\beta^{p-6}\gamma^{1/2}
\label{eq:inflationabc}
\end{align}
The most significant departure from the $q>3$ models is that for layers $l>1$, the number of vertex constraints becomes greater than the number of available spins. This means that the extensive ground state degeneracy found previously is lost. We identify two distinct cases:
\begin{itemize}
    \item {\textit{p} is odd:} The ground state is unique. The geometric constraints imposed by the lattice uniquely fix the state of every spin. The central spin is forced into a specific orientation (e.g., $S_i^z = 1$), and this choice propagates deterministically throughout the entire lattice, leaving no remaining degrees of freedom. An example of this is illustrated in Figure~\ref{fig:q3-tess}$(e)$.
    If the second layer of spins were all pointing down, then the third layer would not be able to find a configuration to reach the ground state.
    \item {\textit{p} is even:} The ground state is four-fold degenerate. This degeneracy arises from two distinct binary choices: One degree of freedom corresponds to the orientational choice of the central spin and a second degree of freedom arises from the choice of the initial spin orientation on the first layer ($k=1$). This is illustrated in Figure~\ref{fig:q3-tess}$(a-d)$.
\end{itemize}

%==========================================================
\lu{\subsection{Tessellations in flat space}
\label{sec:gsdflat}
%============================
\subsubsection{Square lattice (\{4,4\} tessellation)}
Although the Plaquette Ising model on the square lattice is on a flat space, its inflation rule is given by the same formula (\ref{eq:inflationab}) as for hyperbolic lattices. It means one can directly apply Eq.~(\ref{eq:SNa}) to obtain the ground-state entropy,
\begin{align*}
    S = k_B \log{2} \times \left(1 + \sum_{k=1}^{l} 4 \right) = k_B \log{2} \times \left(1 + 4l \right).
\end{align*}
Since a \{4,4\} tessellation of $l$ layers has $(2l +1)^2$ spins, the intensive entropy vanishes in the thermodynamic limit as expected,
\begin{align*}
    s = \lim_{l\to\infty} k_B \log{2} \times \cfrac{1+4l}{(1 + 2l)^2} \to 0.
\end{align*}
On the other hand, since we have $N_\text{surf}=8l$ spins on the boundary, the ground-state entropy scales linearly with the boundary area
\begin{align*}
    \lim_{l\to\infty} \cfrac{S}{N_\text{surf}} = \lim_{l\to\infty}  k_B \log{2} \times \left( \cfrac{1+4l}{8l} \right) = \frac{1}{2}k_B \log{2}.    
\end{align*}
as expected from the discussion of section \ref{sec:efm}.}

\lu{
%============================
\subsubsection{Hexagonal lattice (\{6,3\} tessellation)}
The inflation rule (\ref{eq:inflationabc}) also applies to the $\{6,3\}$ tessellation on flat space
\begin{align*}
\tau: \quad \sigma &\to \beta^6 \\
\beta &\to \gamma^{1/2}\beta\gamma^{1/2} \\   
\gamma &\to \gamma.
\end{align*}
Hence, as for the other $\{p,3\}$ tessellations with even values of $p$, the ground state degeneracy is four fold, as illustrated in Fig.~\ref{fig:6,3-tessellation}. Since $\gamma$ spins are in contact with two spins in the previous layer, their orientations are fixed by the previous layer ($q=3$). And since $\beta$ spins are surrounded by $\gamma$ spins in layer $l$, their orientations are also fixed. It means the finite degeneracy of $\{p,3\}$ tessellations is attached to the local trigonal symmetry of vertices, and not to the negative space curvature.
%%%%%%%%%%%%%%%%%%%
\begin{figure}
    \centering
    \includegraphics[width=1\linewidth]{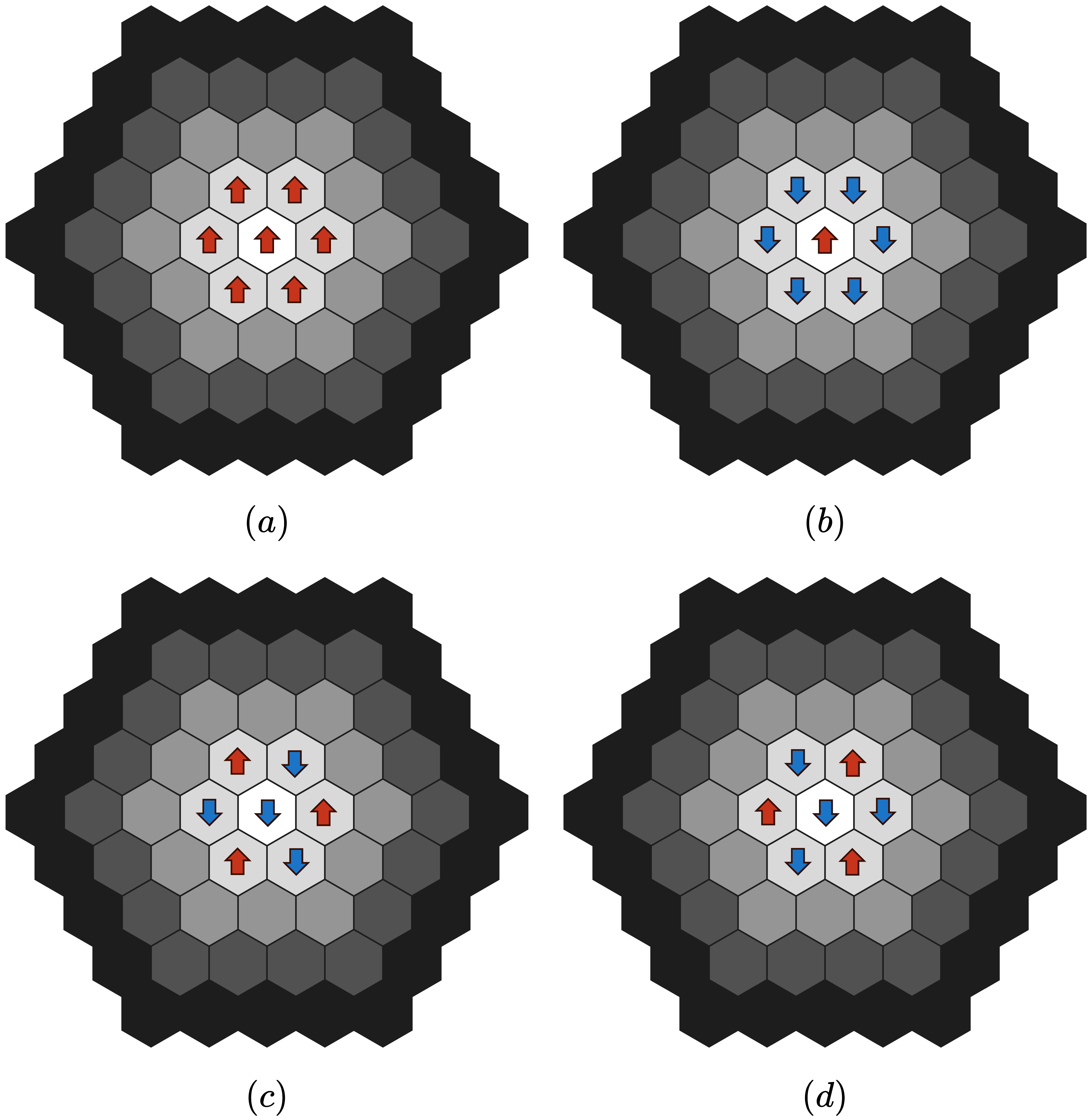}
    \caption{The four ground states of the hexagonal tessellation \{6,3\}, following the same reasoning as in Fig.~\ref{fig:q3-tess} for \{$p,3$\} hyperbolic tessellations with even values of $p$.}
    \label{fig:6,3-tessellation}
\end{figure}
%%%%%%%%%%%%%%%%%%%
%============================
\subsubsection{Triangular lattice (\{3,6\} tessellation)}
Since the inflation formula of Eq.~(\ref{eq:inflationab}) is valid for the $\{3,6\}$ tessellation, one can directly use the formula of the ground-state degeneracy (\ref{eq:sinfty}), which gives
\begin{align*}
    s = \lim_{l\to\infty} \cfrac{S}{N_p} \to \frac{1}{2}k_B \log{2},
\end{align*}
with $R_\infty=1$. The extensive entropy can be understood locally. With six triangles meeting at a vertex the system is under-constrained and allows for loop fluctuations in the ground state, as illustrated in Fig.~\ref{fig:3,6-tessellation}. Similar loop fluctuations are allowed for all  hyperbolic $\{3,q\}$ tessellations.
%%%%%%%%%%%%%%%%%%%
\begin{figure}
    \centering
    \includegraphics[width=1\linewidth]{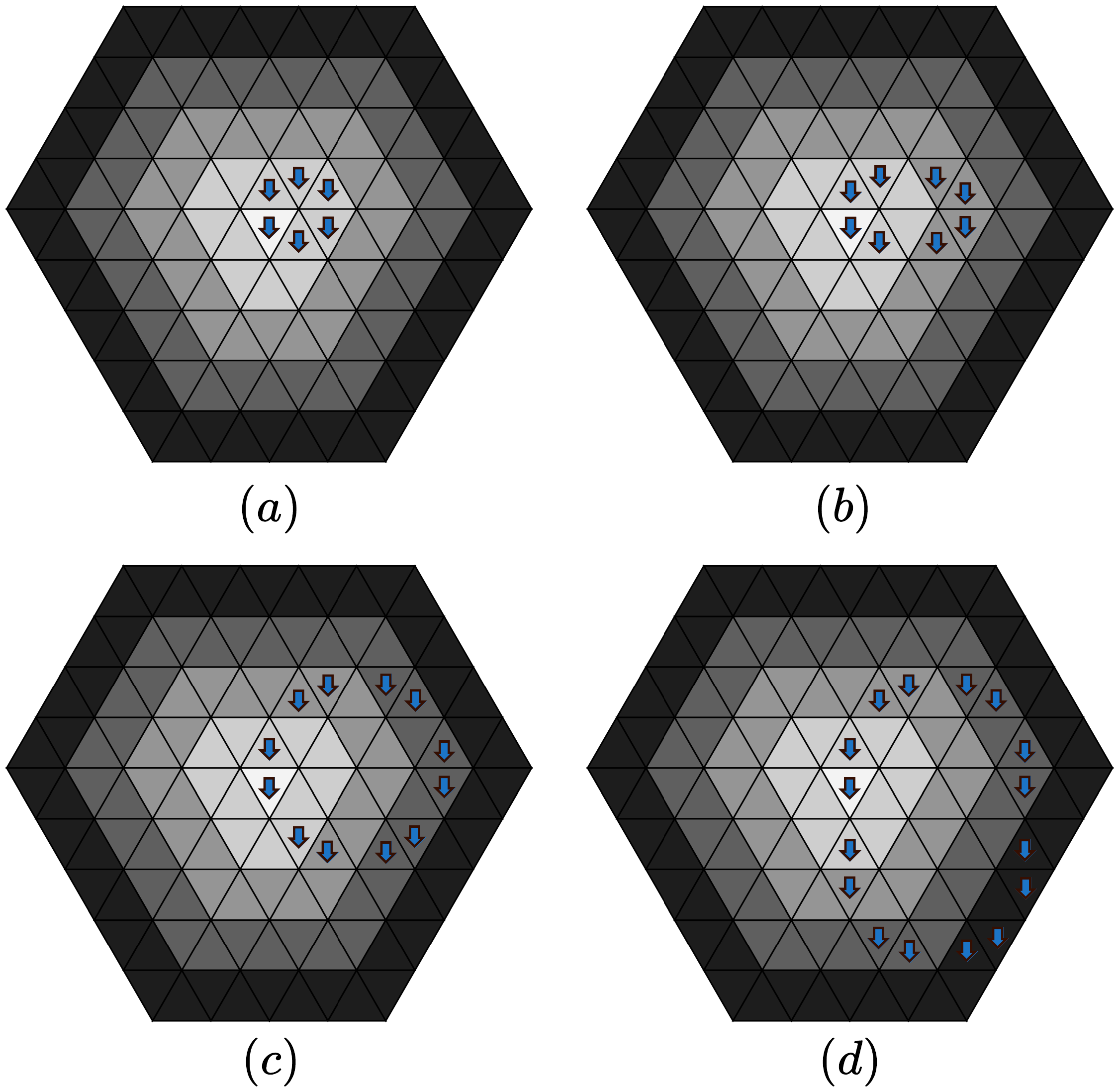}
    \caption{\lu{Example of loops of spin flips that can be formed on the $\{3,6\}$ tessellation, contributing to the extensive ground-state degeneracy. All loops can be built from the elementary hexagon loops.}}
    \label{fig:3,6-tessellation}
\end{figure}
%%%%%%%%%%%%%%%%%%%

\hy{This model is equivalent to the classical limit of the honeycomb color code, 
which realizes a   $\mathbb{Z
}_2$ gapped topological order phase. The local symmetry (flipping a loop) corresponds to exactly the dynamics of the honeycomb color code \cite{PhysRevLett.97.180501}. } 
The \{3,6\} tessellation thus offers the only extensive degeneracy of the HFM in Euclidean space.\\

As a running summary, the degeneracy of the HFM is essentially linked to the connectivity of the lattice; finite for $q=3$, extensive for $q>3$, with the \{4,4\} Plaquette Ising Model as a marginal case with subextensive degeneracy.
}

%==========================================================
\subsection{Subsystem Symmetries}

\begin{figure}
    \centering
    \includegraphics[width=1\linewidth]{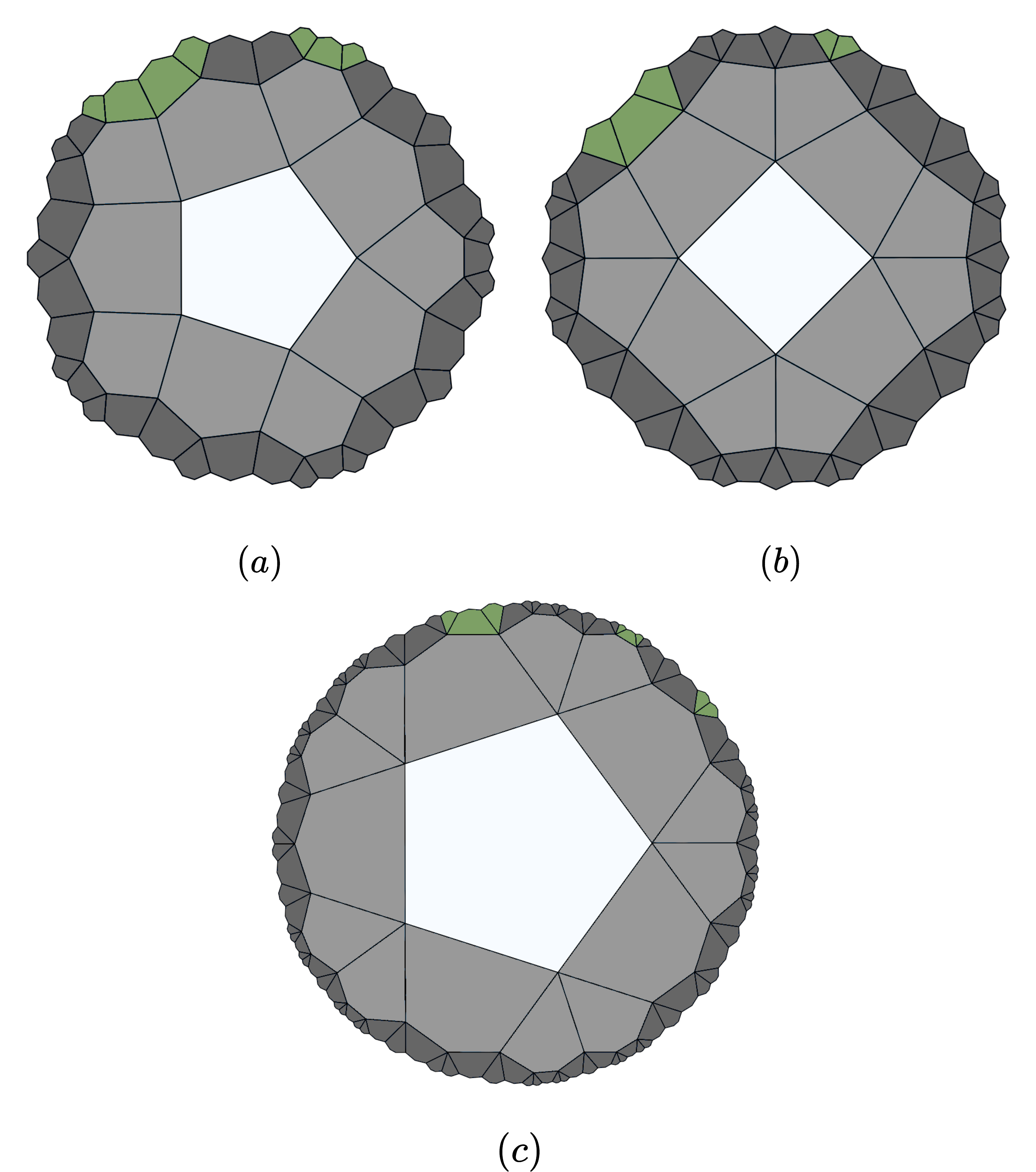}
    \caption{Examples subsystem symmetry that preserve the ground state acting only on the boundary. These sequences of spin flips (green polygons) leave all $\mathcal{O}_v$ operators invariant. $(a)$ On a $\{5,4\}$ lattice, $\alpha\beta\alpha$ and $\alpha\beta\beta\alpha$ sequences are symmetries. $(b)$ On a $\{4,5\}$ lattice, $\alpha\beta\alpha$ and $\alpha\alpha$ sequences are allowed. $(c)$ On a $\{5,5\}$ lattice, the same sequences apply.}
    \label{fig:invariant-op}
\end{figure}

\begin{figure}
    \centering
    \includegraphics[width=1\linewidth]{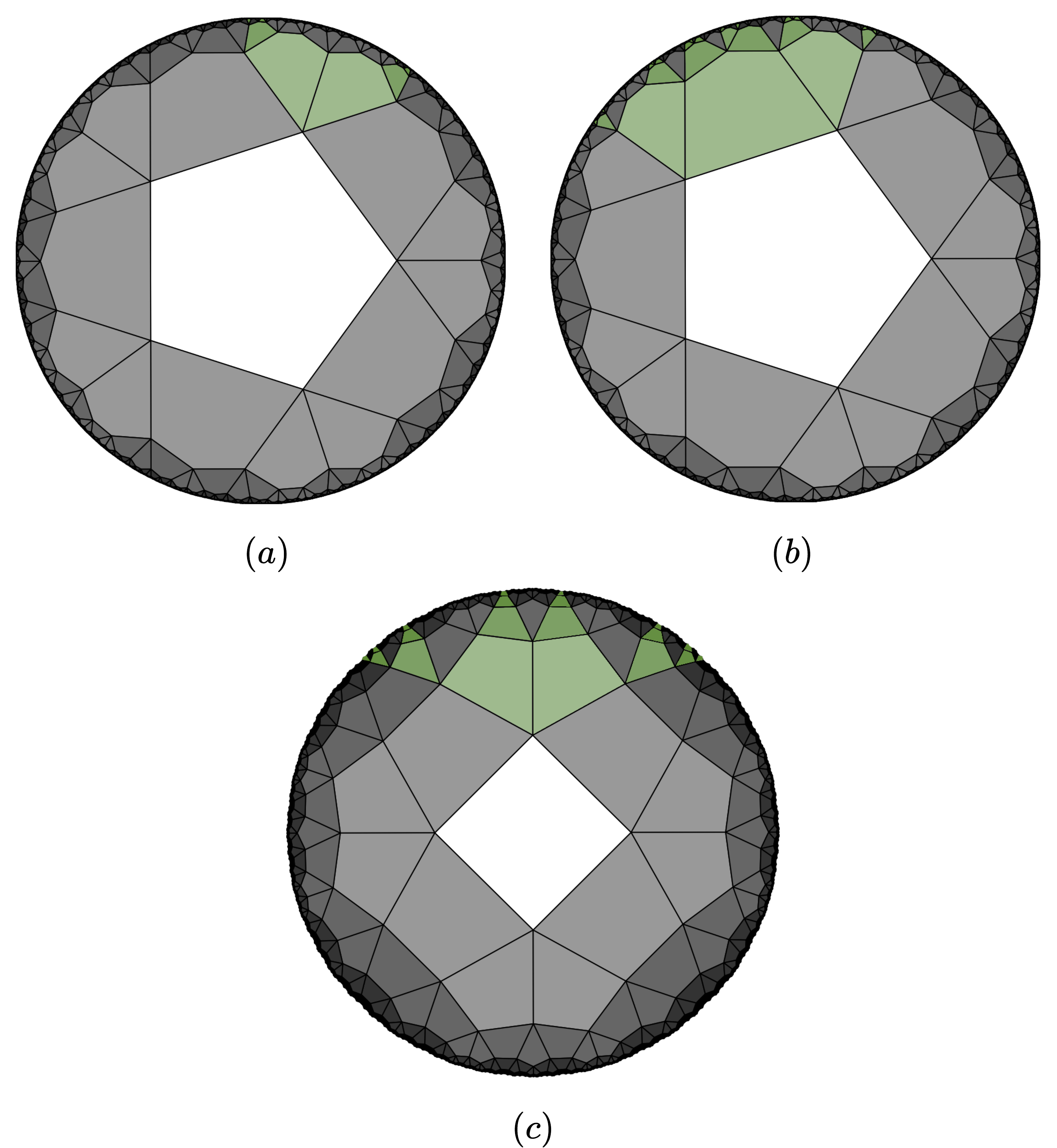}
    \caption{Illustration of the subsystem symmetry operations. An elementary operation in the bulk is propagated towards the boundary, creating self-similar and fractal chains of spin flips that leave the system invariant.}
    \label{fig:bulk-invariance}
\end{figure}

The ground state degeneracy of the HFM is a direct consequence of the subsystem symmetries of the system. These symmetries are \textit{invariance operations} that map one state to another while conserving the energy.
%In particular they enable to move around the ground-state manifold. 
The simplest symmetric operations are those on the boundary, notably the sequence of spin flips on adjacent spins ${\alpha\beta\alpha}$, ${\alpha\beta\beta\alpha}$ or ${\alpha\alpha}$. These sequences of spin flips, which we refer to as \textit{``elementary''}, preserve the sign of operators $\mathcal{O}_v$ on the boundary as illustrated in Figure~\ref{fig:invariant-op}. 

However, applying one of the elementary operations in the bulk creates excitations on the next layer. In order to erase the excitations, one needs to flip a fractal tree of spins all the way to the boundary, constructing this way the bulk subsystem symmetry operations of the system. These symmetries are subsystem topological in the sense that they require an extensive number of spin flips, and exhibit self-similar patterns as shown in Figure~\ref{fig:bulk-invariance}.

\lu{In particular, for all \{$3,q$\} tessellations, the subsystem symmetry is extensive; any ground state is connected to an extensive number of other ground states via closed loop moves as illustrated for the \{$3,6$\} lattice in Fig.~\ref{fig:3,6-tessellation}. This will have important consequences for the Rindler reconstruction discussed in the next chapter.}
\section{Rindler Reconstruction}\label{sec:rindler_reconstruction}
A central idea of holography is the ability to reconstruct bulk information from boundary data. In the context of the HFM, this is realized through the \textit{Rindler reconstruction}, which is a discrete analogue to the entanglement wedge reconstruction in the AdS/CFT correspondence. This process demonstrates how a region of the bulk spin configuration can be uniquely determined from a boundary subregion $A$. The bulk region that can be reconstructed is bounded by the \textit{minimal wedge} $\Gamma$, which is defined as the shortest path connecting the endpoints of $A$.

\begin{figure}
    \centering
    \includegraphics[width=1\linewidth]{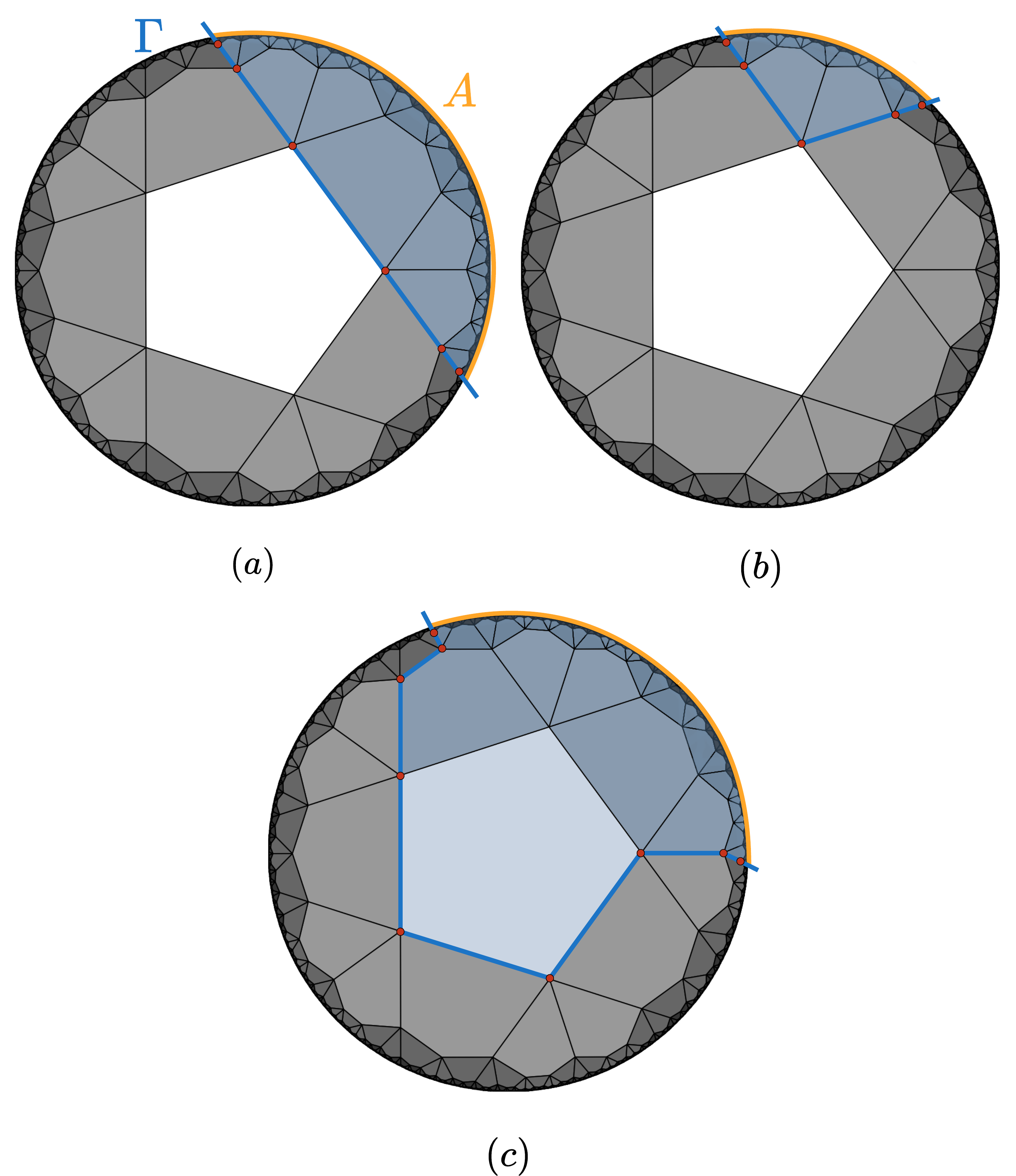}
    \caption{Illustration of the minimal wedge on the $\{5,5\}$ tessellation. The mutual information $I(A,B)$ is directly proportional to the number of vertices (represented by red dots), $N_{v_\Gamma}$, located on the minimal wedge $\Gamma$ separating the two subregions. Each such vertex represents a broken constraint, contributing one bit to the mutual information. The geometric length of the boundary segment is given by $|\Gamma| = N_{v_\Gamma} + 1$.}
    \label{fig:mutual_info}
\end{figure}

The reconstruction begins by assigning values to all spins on the boundary subregion $A$. For each operator $\mathcal{O}_v=1$ adjacent to the boundary, if $q-1$ of its spins are known then the value of the remaining bulk spin is uniquely determined. By applying this procedure recursively,
we can reconstruct the bulk spins inward until all constraints are exhausted. The reconstruction process halts when all remaining vertex operators involve two or more undetermined spins, making further reconstruction impossible. The reconstructed bulk region is enclosed by the minimal wedge, which is illustrated in~\ref{fig:mutual_info}.

\hy{Essentially, the model needs to have sub-extensive subsystem symmetry to make Rindler reconstruction work. That is, the symmetry acts on the boundary and also extends in to the bulk infinitely deep in the thermodynamical limit. So whatever action applied to the boundary must be extended to  the bulk within the ground state sector. 
Therefore, for $p=3$, the local symmetry \lu{which allows for an extensive number of loop moves does not support the Rindler reconstruction. This reconstruction does, however, work} generally for $p\ge 4$.}

%Relation to quantum error correction:  bulk states are protected against errors since we need to erase a large region to prevent their reconstruction while boundary states are easily removed...

%Suppose a portion of the total boundary states are corrupted. We then apply the Rindler Reconstruction on the disconnected subregions in order to recover the bulk states of the union of these disconnected subregions. If we can reconstruct boundary states from the bulk then we could recover the corrupted boundary subregions.

%Given a bulk region, how much boundary states can we reconstruct, if any?

\section{Mutual information}\label{sec:mutual}
Having established the Rindler reconstruction, we now turn to the relationship between boundary entanglement and bulk geometry. In AdS/CFT, this is captured by the Ryu-Takayanagi (RT) formula, which relates entanglement entropy  between two complimentary boundary subregions to the area of minimal covering surfaces in the bulk~\cite{Ryu2006PhysRevLett,Ryu2006JHEP}. In this section, we demonstrate that the HFM on arbitrary $\{p,q\}$ tessellation provides an analogue of this principle, where the mutual information between boundary subregions is determined by the length of minimal wedge.

Mutual information is a measure in information theory that quantifies the shared information between two systems. For two subsystems, $A$ and $B$, it is defined in terms of their entropies as:
\begin{eqnarray}
    I(A,B) = S(A) + S(B) - S(A \cup B)
\end{eqnarray}
where $S(A)$ is the entropy of subsystem $A$, and $S(A \cup B)$ is the entropy of their union. It has been pointed out in~\cite{Yan2019PhysRevBfracton1,Yan2019PhysRevBfracton2} that the mutual information serves as a classical analog of the entanglement entropy between a bipartition of a quantum system. 

\subsection{Mutual Information for Connected Subregions}

Let us first consider a bipartition of the boundary into two  subregions, $A$ and $B$. In order to compute the mutual information $I(A,B)$, we use the vertex representation established in Section~\ref{sec:gsd}. We first classify the spins and vertices of the tessellation with respect to their subregions. The total number of spins $N_p$ and vertices $N_v$ can be partitioned as follows:
\begin{gather}
    N_p = N_{p_A} + N_{p_B} \label{eq:mutual_poly} \\
    N_v = N_{v_A} + N_{v_B} + N_{v_\Gamma} \label{eq:mutual_vertex}
\end{gather}
where $N_{p_A}$ and $N_{p_B}$ are the number of spins contained entirely within the reconstruction wedge of subregions $A$ and $B$, respectively. Here, we assumed that such clear separation can be made, i.e., that each spin  belongs to one and only one of the three following cases: the reconstruction wedge of A,  the reconstruction wedge of B, or to the boundary of the two reconstruction wedges. There is no overlap between the two reconstruction wedges, and the two wedges and their boundary cover the entire lattice. This is not always the case due to lattice discretization, but in the thermodynamics limit, the deviation is finite and small \cite{Yan2019PhysRevBfracton1}. Similarly, $N_{v_A}$ and $N_{v_B}$ are the number of vertices fully interior to each subregion, while $N_{v_\Gamma}$ is the number of vertices lying on the boundary $\Gamma$ that separates them.

The GSD for a subregion $A$ is determined by the number of its unconstrained degrees of freedom, $\Omega(A) = 2^{(N_{p_A} - N_{v_A})}$. The corresponding entropy is:
\begin{gather*}
S(A) = k_B\log 2 \times (N_{p_A} - N_{v_A})
\end{gather*}
with an analogous expression for the subsystem $B$. The joint entropy of the combined system $A \cup B$ is simply the total entropy of the entire lattice:
\begin{gather*}
    S(A \cup B) = k_B\log 2 \times (N_p - N_v)
\end{gather*}

Substituting these expressions into the definition of mutual information yields:
\begin{equation}
\begin{split}
    I(A,B) = &k_B\log 2 \times [(N_{p_A} - N_{v_A}) \\
    &+ (N_{p_B} - N_{v_B}) - (N_p - N_v) ]
\end{split}
\end{equation}
By applying the conservation relations from Eqs.~\eqref{eq:mutual_poly} and \eqref{eq:mutual_vertex}, we arrive at the simple result:
\begin{gather}\label{eq:mutual_result}
    I(A,B) = k_B\log 2 \times N_{v_\Gamma}
\end{gather}

The interpretation of this result lies in the structure of the HFM. For each subregion, spins located next to the boundary $\Gamma$ between $A$ and $B$ are no longer subject to the constraints on this boundary, since part of their neighboring spins lie in the opposite subregion. Each vertex on the boundary $\Gamma$ corresponds to a broken constraint which gives an unconstrained DOF. This unconstrained DOF contributes exactly $k_B \log 2$ to the mutual information.

The geometric length of the boundary segment $\Gamma$ is given by $|\Gamma| = N_{v_\Gamma} + 1$, where the length of the boundary corresponds to the number of edges between consecutive vertices. The mutual information for a connected bipartition can therefore be expressed as:
\begin{gather}\label{eq:RT_formula}
    I(A,B) = k_B\log 2 \times (|\Gamma| - 1) \approx k_B\log 2 \times |\Gamma|
\end{gather}
This is a discrete realization of the Ryu-Takayanagi formula, establishing a link between the information shared across a boundary subregion and the geometry of that boundary.

\subsection{Mutual Information for Disconnected Subregions}

The framework presented above can be generalized to scenarios where the boundary is partitioned into multiple disconnected subregions. In such cases, the calculation remains the same, but corrections must be taken into account. Consider a configuration where subregion $A$ consists of $n_\Gamma$ disconnected components. Each boundary segment $\Gamma_i$ separating these components from their complement contributes independently to the mutual information.

For a total of $n_\Gamma$ disconnected boundary segments, the Ryu-Takayanagi formula is modified to:
\begin{gather}\label{eq:RT_disconnected}
    I(A,B) = k_B\log 2 \times \left( \sum_{i=1}^{n_\Gamma}|\Gamma_i| - n_\Gamma \right)
\end{gather}
where $N_{v_\Gamma} = \sum_{i=1}^{n_\Gamma} N_{v_{\Gamma_i}}$. The correction term $n_\Gamma$ arises from the discretization of the hyperbolic lattice. In our counting procedure, the geometric length $|\Gamma_i|$ of each segment includes its endpoints, leading to an over-counting of vertices when the segments are disconnected. The correction term accounts for this, ensuring that each disconnected boundary contributes an amount proportional to its length minus one which is consistent with the connected case.
\section{Black Hole Entropy}\label{sec:BH}

\begin{figure}
    \centering
    \includegraphics[width=1\linewidth]{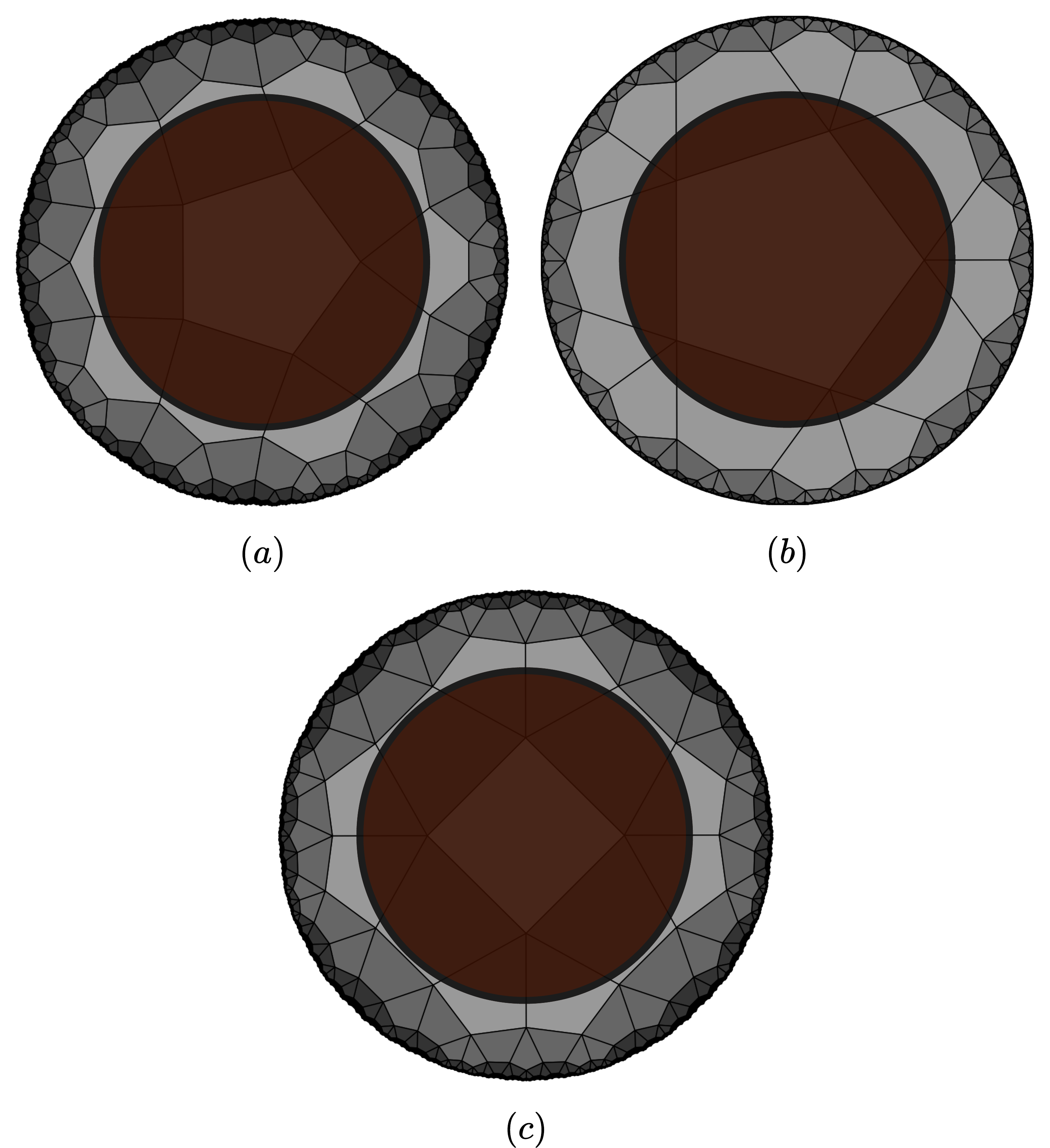}
    \caption{Construction of a naive black hole on $(a)$ $\{5,4\}$, $(b)$ $\{5,5\}$ and $(c)$ $\{4,5\}$ hyperbolic tessellations. The underlying geometry of the lattice is preserved, but all spins and vertices inside the solid black line (the horizon) are removed from the system. Consequently, external observers can only access the degrees of freedom outside the horizon.}
    \label{fig:black-hole}
\end{figure}

Building upon the holographic properties discussed in the preceding sections, we now turn to the physics of black holes. We define a black hole by selecting a closed convex region within the lattice and designate its perimeter as the event horizon. All spins and vertices inside this region are considered "hidden" to an external observer, while the rest of the lattice remains unchanged (i.e., one can think of this as removing all the spins and vertex terms in the Hamiltonian). The introduction of a black hole alters the structure of the system, leading to an increase in the ground state degeneracy and consequently the entropy. We define the black hole entropy $S_{\text{BH}}^i$ as the change in the total entropy of the system upon introducing the BH.

To derive an analytical expression for the black hole entropy and examine its scaling properties, we now switch to the polygon representation. Let the black hole be formed in a highly symmetric way, by removing all layers up to layer $i-1$ around a selected center. The entropy of the unperturbed system with $l$ layers is given by:
\begin{gather*}
    S = k_B\log{2} \left(\sum_{k=1}^{l} N^k_{\alpha} + 1 \right)
\end{gather*}
In the perturbed system the $\beta$-spins at layer $i$ are now free to fluctuate. The entropy of the perturbed system is thus:
\begin{gather*}
    S_{\text{perturbed}}^i=  k_B\log{2} \left(N^i_{\beta} + \sum_{k=i}^{l} N^k_{\alpha} + 1 \right)
\end{gather*}
The change in entropy which is the black hole entropy $S_{\text{BH}}^i$ is the difference between these two quantities:
\begin{gather*}
    S_{\text{BH}}^i = k_B\log{2} \left(N^i_{\beta} - \sum_{k=1}^{i-1} N^k_{\alpha} \right)
\end{gather*}
The quantities $N^i_{\beta}$ and $\sum N^k_{\alpha}$ can be expressed analytically using Eqs.~(\ref{eq:naal},\ref{eq:nbbl})
\begin{eqnarray}
    \sum_{k=1}^{i-1} N^k_{\alpha} = A_+\frac{\lambda_+^{i-1}-1}{\lambda_+ -1} + A_-\frac{\lambda_-^{i-1}-1}{\lambda_- -1}\label{eq:naalsum}
\end{eqnarray}
which gives the exact expression for the black hole entropy:
\begin{widetext}
\begin{eqnarray}
    S_{\text{BH}}^i = k_B\log{2} \left[ \left( B_+ - \frac{A_+}{\lambda_+ -1}  \right)\lambda_+^{i-1} 
    + \left( B_- - \frac{A_-}{\lambda_- -1}  \right)\lambda_-^{i-1} + \left(\frac{A_+}{\lambda_+ -1} + \frac{A_-}{\lambda_- -1}\right) \right]
\end{eqnarray}
\end{widetext}
where the coefficients $A_\pm, B_\pm, \lambda_\pm$ depend on the lattice parameters $\{p,q\}$ and are expressed in Sec.~\ref{sec:inflationmatrix}. The black hole horizon's area corresponds to the number of vertices located on the horizon $N_v^i = N_b^i$ (see Eq.~(\ref{eq:Npabv})). In the limit where the black hole becomes very large ($i \to \infty$), the terms associated with the largest eigenvalue $\lambda_+$ dominate. By calculating the entropy per horizon area, we show that it converges to a non-zero constant:
\begin{gather*}
    \lim_{i \to \infty}\frac{S_{\text{BH}}^i}{N_v^i} = k_B\log{2} \times \left(1 - \frac{1}{R_\infty(\lambda_+ -1)}\right) 
\end{gather*}
This result confirms that the black hole entropy scales linearly with the horizon's area, providing the discrete analogue of the Bekenstein-Hawking formula~\cite{Hawking_1975}. 

\hy{Finally, we note that in this section we considered the ``black hole'' as a toy-model analogue rather than a gravitational black hole in the sense of general relativity. Our construction follows the spirit of holographic tensor-network models defined on discretized hyperbolic lattices~\cite{Pastawski2015}, where a black hole is represented by removing a finite region in the internal of  the lattice and introducing additional degrees of freedom associated with the resulting boundary. Because the lattice geometry is a regular $\{p,q\}$ tessellation of the hyperbolic plane, the curvature remains uniform everywhere and no metric singularity is present. The effective ``horizon'' therefore arises simply from the cutoff of a finite region of the lattice. The hyperbolic tiling terminate at this boundary and do not probe the interior region, with dangling degrees of freedom at the boundary, reproducing the black hole ``soft hair'' degrees of freedom. The full geometric and dynamical properties of gravitational black holes are not reproduced.}

\section{Fracton Excitations}\label{sec:fractons}

\subsection{Propagation rules for fractons}
\label{app:fracton_rules}

\begin{figure}
    \centering
    \includegraphics[width=1\linewidth]{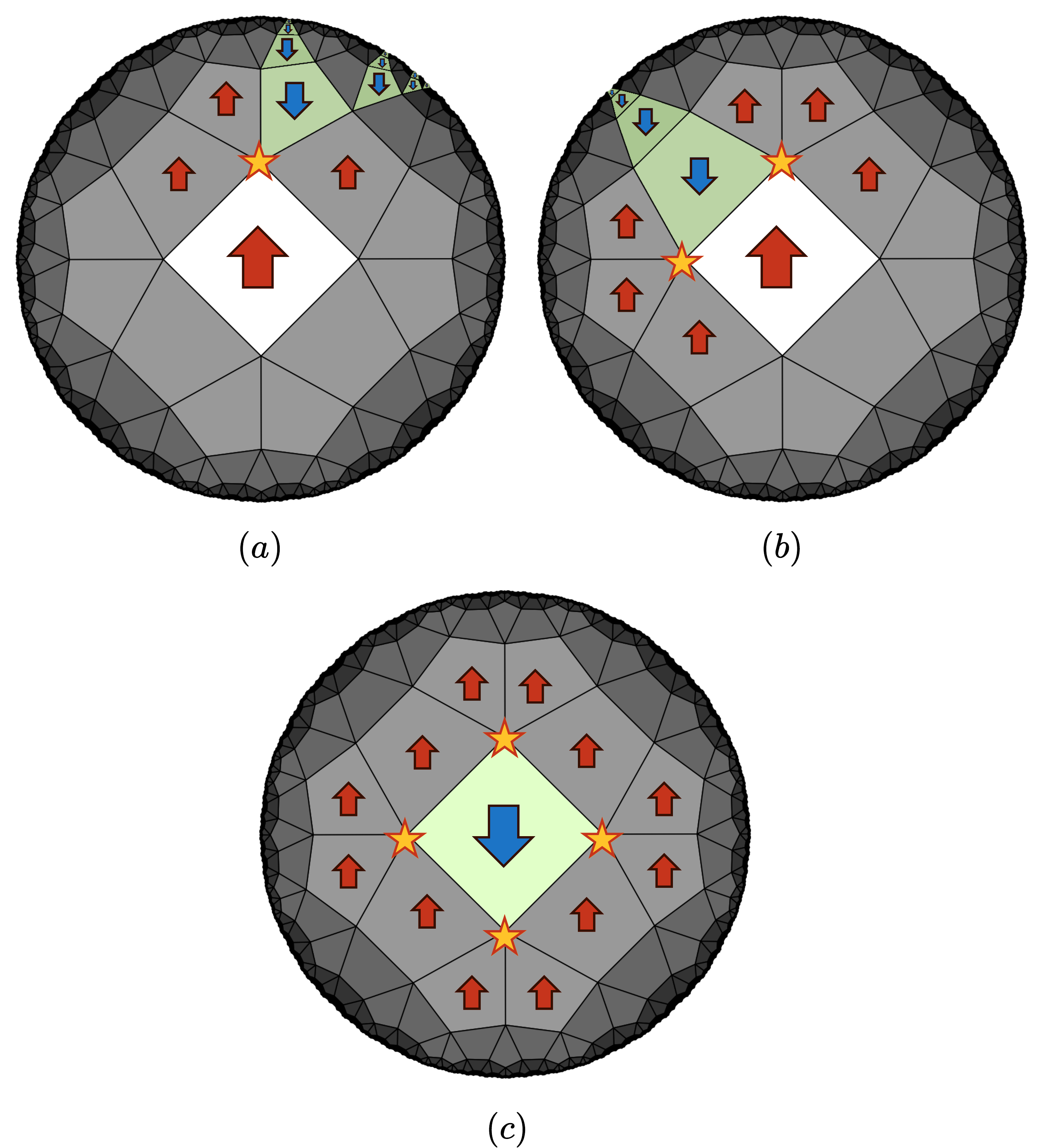}
    \caption{Fracton excitations on the $\{4,5\}$ tessellation. Stars denote vertex violations ($\mathcal{O}_v = -1$), while green polygons show the spins that must be flipped to preserve the ground state for the other vertices. $(a)$ Creating a single, immobile fracton requires flipping an initial $\alpha$-spin, followed by a fractal-like sequence of spins on subsequent layers. $(b)$ A mobile, bound fracton pair can be created by flipping a line of $\beta$-spins. $(c)$ Flipping a single spin in the bulk creates a four-fracton bound state, violating the four adjacent constraints.}
    \label{fig:fracton_excitations}
\end{figure}

We have so far focused on the ground state of the HFM. But one of its main properties is also the existence of fracton excitations. Flipping a single spin in the bulk of a $\{p,q\}$ lattice creates a $p$-fracton bound state (Fig.~\ref{fig:fracton_excitations}(c)). In order to create a single fracton excitation we need to flip the sign of a single $O_v$ constraint while keeping all others intact. How to achieve this differs between $\alpha$ and $\beta$ spins. Flipping one $\alpha$ spin creates one fracton on layer $k-1$ and $(p-1)$ fractons on layer $k$, while flipping one $\beta$ spin creates two fractons on layer $k-1$ and $(p-2)$ fractons on layer $k$. In order to preserve the fracton(s) on layer $k-1$ only, we need to erase the excitations of layer $k$. To do so, we flip spins on layer $k$, which ``pushes'' the fractons from layer $k$ to $k+1$. Because of the negative curvature of hyperbolic lattices, the number of fractons generally increases in the process. Repeating the procedure keeps pushing the fractons to higher layers, until they finally vanish on the boundary, given proper open boundary condition.%(similarly to the invariance operations mentioned before). 
One is then left with either a single or a pair of fractons, depending on the nature of the initial spin flip: respectively an $\alpha$ spin in Fig~\ref{fig:fracton_excitations}a) or a $\beta$ spin in Fig~\ref{fig:fracton_excitations}b). The resulting fracton excitation is localized in the system because it cannot move at zero-energy cost without flipping an extensive number of spins. Moreover, it is topological in the thermodynamic limit $l\to\infty$ because no finite sequence of spin flips can create a single excitation. 

%Fractal self similar operators appear for q > 4, is there a link with Haah's code?

In general, there is no unique  or highly symmetric way to choose the extensive fractal chain of spins to flip. But for a given tesselation $\{p,q\}$, there is a certain propagation rule given in Table \ref{tab:prop_rules} that minimizes the number of fractons created at each step.
This is particularly important, since it shows the minimal energy barrier the system needs to overcome to push a fracton away to the boundary. 

%\begin{itemize}
%    \item Flipping an $\alpha$-spin introduces $(p-4)$ flips on $\beta$-spins on the next layer, distributed symmetrically around the location of the initial flip. If $p$ is even, an additional $\alpha$-spin within the central $\alpha^t$ sequence must also be flipped.
%    \item Flipping a $\beta$-spin introduces $(p-5)$ flips on $\beta$-spins on the next layer, also distributed symmetrically. If $p$ is odd, an additional $\alpha$-spin in the central $\alpha^t$ sequence must be flipped.
%    \item {Boundary:} The leftmost and rightmost $\beta$-spins in the propagation sequence are always the first to be flipped.
%    \item {Spacing:} There is always a sequence of unflipped polygons corresponding to $\alpha^t\beta\alpha^t$ between the flipped $\beta$-spins. The only exception to this rule occurs when an extra $\alpha$-spin is flipped from within this central sequence.\\
%\end{itemize}

\subsection{Scaling growth of fractons}
\label{app:fracton-growth}

\begin{figure}
    \centering
    \includegraphics[width=1\linewidth]{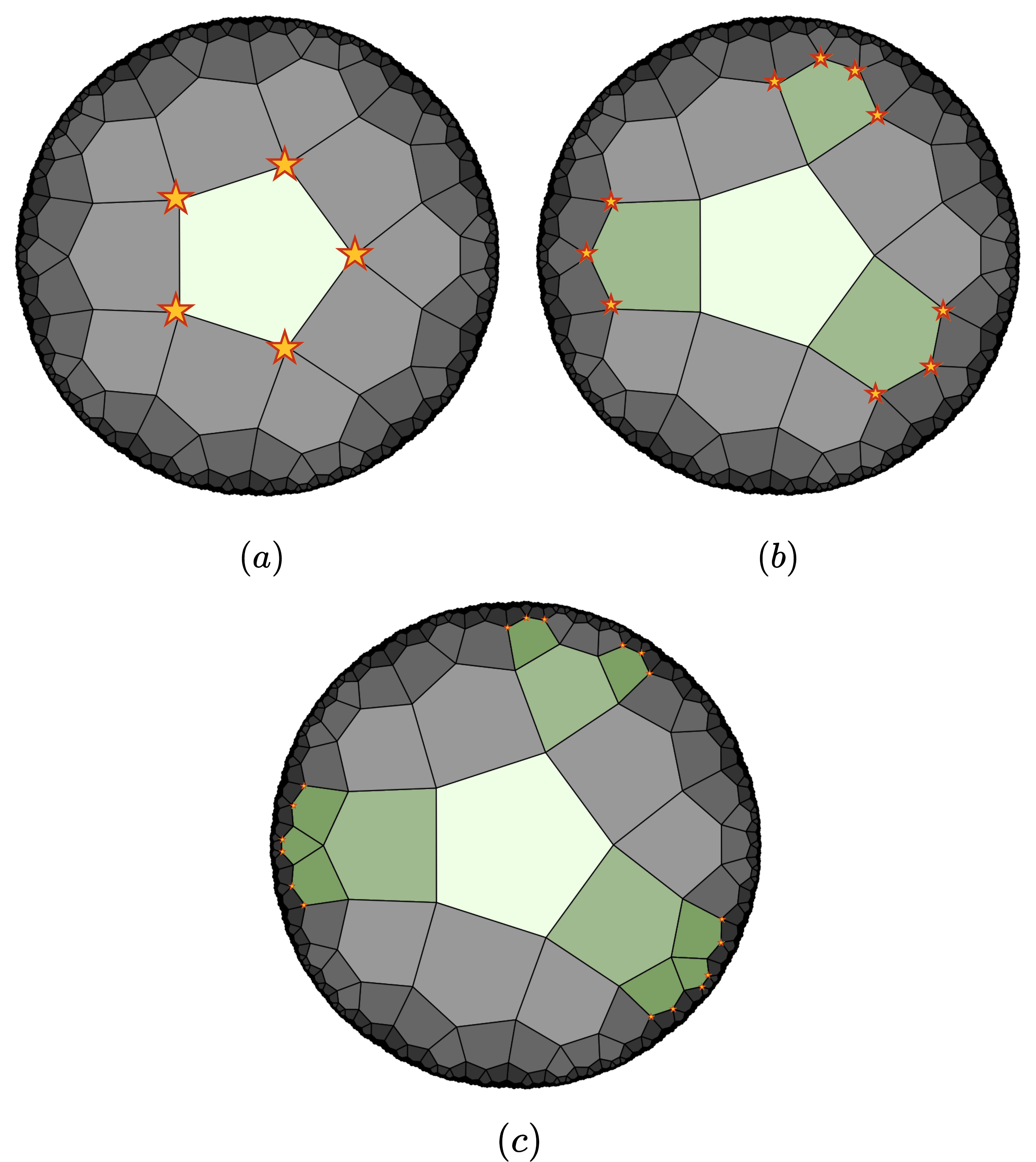}
    \caption{Propagation of fracton excitations on the $\{5,4\}$ tessellation. Stars denote vertex violations ($\mathcal{O}_v = -1$), while green polygons show the flipped spins. $(a)$ Flipping a single spin creates 5 fractons. $(b)$ Flipping two $\beta$-spins and an $\alpha$-spin pushes the fractons to the next layer. $(c)$ Repeating this procedure, the fractons are pushed further until they vanish at the boundary. The negative curvature of hyperbolic lattices induces an increase in the number of fractons at each step.}
    \label{fig:fracton_propagation}
\end{figure}

From the propagation rules of Table \ref{tab:prop_rules}, it is straightforward to compute how the minimum number of fractons $N_{\text{frac}}(p,q,k)$ for a given tesselation $\{p,q\}$ grows from layer $k$ to $k+1$ when being pushed outward. For convenience, let us flip the central spin ($k=0$) which creates $p$ fractons, and then push the fractons outward layer after layer as illustrated in Fig~\ref{fig:fracton_propagation}. $N_{\text{frac}}(p,q,k)$ is the number of  fracton excitations as it propagates outward to layer $k$. Once we reach the boundary, there is no more fracton, and we have performed a subsystem symmetry operation, going from one ground state to another. For any number $q$ of spins around a vertex at layer $k$, we only need to flip two spins to push the fracton to the next layer $k+1$; this is independent of how many spins are around the vertex and we thus have $N_{\text{frac}}(p,q,k)$ independent of $q$ for most tesselations. We show that for all tessellations $\{p,q\}$ with $p>4$ and $q>3$, the number of fracton excitations on the HFM grows exponentially with layer $k$ as fractons are pushed to the boundary. The hyperbolic lattice with smallest Schl\"afli symbol $\{5,4\}$ is the only one with a non-monotonic growth that depends on the parity of $k$.

%\subsubsection{Tessellations with Even \textit{p}}
For a generic tessellation where the polygons have an even number of sides $p$, the number of fractons grows exponentially with the layer number $k$:
\begin{eqnarray}
N_{\text{frac}}(p\in 2\mathbb{Z},q,k) = p \left( \frac{p-2}{2} \right)^{k-1}
\end{eqnarray}

A special case is the  square tessellations, whose   number of fractons is constant and equal to:
\begin{eqnarray}
N_{\text{frac}}(4,q,k) = 4 
\end{eqnarray}
The fracton pair has a sub-dimensional dynamics as it can move along a chain of $\beta$ spins at no energy cost (Fig~\ref{fig:fracton_excitations}b).
%\subsubsection{Triangular tesselation $\{3,q\}$}

Next, let us examine the cases when $p$ is odd. 
For the triangular tessellations, \lu{including the \{3,6\} Euclidean tessellation}, the number of fractons remains constant after the initial layer:
\begin{eqnarray}
N_{\text{frac}}(3,q,k) = 3 
\end{eqnarray}

\lu{As a consequence, the two Euclidean tessellations with (sub-)extensive ground-state degeneracy preserve a constant number of fractons as they are expelled outwards.}

%\subsubsection{Square tesselation $\{4,q\}$}

%\subsubsection{Pentagonal tesselation $\{5,4\}$}
The $\{5,4\}$  (pentagon) tessellation displays a non-monotonic growth pattern that depends on the parity of the layer number $k$:
\begin{eqnarray}
        N_{\text{frac}}(5,4,k) = 
        \begin{cases}
        5  \quad \text{for $k=1$}\\
        10 \times 3^{(k-2)/2}  \quad \text{for even $k$}\\
        2 \times 3^{(k+1)/2} \quad \text{for odd $k>1$}
        \end{cases}
    \end{eqnarray}

%\subsubsection{Pentagonal tesselation $\{5,q>4\}$}
For pentagonal tessellations with a higher coordination number, the growth is simply exponential:
\begin{eqnarray}
N_{\text{frac}}(5,q>4,k) = 5 \times 2^{k-1}
\end{eqnarray}

%\subsubsection{Heptagonal tesselation \{7,q\}}
For heptagonal tessellations, the geometric sequence has a different common ratio:
\begin{eqnarray}
N_{\text{frac}}(7,q,k) = 7 \times 3^{k-1}
\end{eqnarray}
For tessellations with a large number of odd sides $p \geq 9$ and $q > 3$, the minimal number of fractons follows a general set of rules, which can be inferred from the propagation rules of \{9,q\} given in Table~\ref{tab:prop_rules}, however, no general formula has been found for these tessellations. For example, to derive a formula for the \{9,4\} tessellation we proceed as follows: We flip the central spin which creates nine fractons. To push the fractons to the next layer we need to flip four $\beta$ spins and one $\alpha$ spin. To go one layer further, for each $\beta$ spin we need to flip two $\beta$ spins and one cluster of $\beta\alpha\beta$ spins, and for the $\alpha$ spin we flip four $\beta$ spins. On the following layer, we need to flip nine $\beta$ spins for the $\beta\alpha\beta$ cluster. We deduce the following recurrence relation for layers $k>1$ and write it in matrix form:
\begin{eqnarray}
\begin{split}
\begin{pmatrix}
     N_{\alpha}^{k}  \\ 
     N_{\beta}^{k}  \\ 
     N_{\beta\alpha\beta}^{k}  \\ 
\end{pmatrix} =
&
\begin{pmatrix}
     0 & 0 & 0 \\ 
     4 & 2 & 9 \\ 
     0 & 1 & 0 \\ 
\end{pmatrix}^{k-2}
\begin{pmatrix}
     N_{\alpha}^{2} \\ 
     N_{\beta}^{2} \\ 
     N_{\beta\alpha\beta}^{2} \\ 
\end{pmatrix}
\\\\
&
\begin{pmatrix}
     N_{\alpha}^{2} \\ 
     N_{\beta}^{2} \\ 
     N_{\beta\alpha\beta}^{2} \\ 
\end{pmatrix} = \begin{pmatrix}
     1 \\ 
     4 \\ 
     0 \\ 
\end{pmatrix}
\end{split}
\end{eqnarray}
After computing $N_{\alpha}^{k}$, $N_{\beta}^{k}$ and $N_{\beta\alpha\beta}^{k}$ we multiply each term by its corresponding number of fractons and sum them to get the total number of fractons $N_{\text{frac}}(9,4,k)$ at layer $k$. A flipped $\alpha$ spin corresponds to eight fractons, a $\beta$ spin to seven fractons and a $\beta\alpha\beta$ cluster to twenty fractons.

\begin{table*}[t]
    \centering
    \caption{
    Propagation rules of fracton excitations for $\{p,q\}$ tessellations. To concisely represent the propagation rules, we used the following notations. An underlined symbol, such as \underline{$\alpha$} or \underline{$\beta$}, denotes a flipped spin associated with that polygon type. The exponents are $r = (q-4)/2$ and $t = q-3$. An underlined symbol with an exponent, such as \underline{$\alpha^r$}, indicates that only one of the $\alpha$-spins within that sequence of $r$ (or $t$) spins is flipped. The choice of which spin to flip corresponds to different but symmetrically equivalent propagation patterns. The $\{5,4\}$ tesselation is special because $r=0$, and there is thus no spin to flip in the sequence \underline{$\alpha^r$}.  
    }
    \label{tab:prop_rules}
    \renewcommand{\arraystretch}{1.5}
    \begin{tabular}{@{} l c >{$}l<{$} >{$}l<{$} >{$}l<{$} @{}}
        \toprule
        & \textbf{Tessellation} & \multicolumn{1}{c}{\textbf{Rule for $\underline{\alpha}$}} & \multicolumn{1}{c}{\textbf{Rule for $\underline{\beta}$}} &
        N_\text{frac}(p,q,k)\\
        \midrule
        
         & \{3,q\} & \underline{\alpha} \to \alpha^r \underline{\beta} \alpha^r & \underline{\beta}  \to \underline{\alpha^{-1}} & N_{\text{frac}}(3,q,k) = 3 
\\\midrule
        \addlinespace
        
         & \{4,q\} & \begin{array}{@{}l@{}} \underline{\alpha} \to \alpha^r \underline{\beta} \alpha^t \beta \underline{\alpha^r} \\ \text{or} \\ \underline{\alpha} \to \underline{\alpha^r} \beta \alpha^t \underline{\beta} \alpha^r \end{array} & \underline{\beta} \to \alpha^r \underline{\beta} \alpha^r  & N_{\text{frac}}(4,q,k) = 4\times  2^{k-1} \\\midrule
        \addlinespace

        & \{5,4\} & \underline{\alpha} \to \alpha^r \underline{\beta} \alpha^t \beta \alpha^t \underline{\beta} \alpha^r & \underline{\beta} \to \underline{\beta\alpha\beta} &
        N_{\text{frac}}(5,4,k) = 
        \begin{cases}
        10 \times 3^{(k-2)/2}  \quad \text{for even $k$}\\
        2 \times 3^{(k+1)/2} \quad \text{for odd $k$}
        \end{cases}\\\midrule
        \addlinespace
        
        & \{5,q$>$4\} & \underline{\alpha} \to \alpha^r \underline{\beta} \alpha^t \beta \alpha^t \underline{\beta} \alpha^r & \begin{array}{@{}l@{}} \underline{\beta} \to \alpha^r \underline{\beta} \alpha^t \beta \underline{\alpha^r} \\ \text{or} \\ \underline{\beta} \to \underline{\alpha^r} \beta \alpha^t \underline{\beta} \alpha^r \end{array} &
        N_{\text{frac}}(5,q>4,k) = 5 \times 2^{k-1}
\\\midrule
        \addlinespace
        
        &\{6,q\} & \underline{\alpha} \to \alpha^r \underline{\beta} \alpha^t \beta \underline{\alpha^t} \beta \alpha^t \underline{\beta} \alpha^r & \underline{\beta} \to \alpha^r \underline{\beta} \alpha^t \beta \alpha^t \underline{\beta} \alpha^r & N_{\text{frac}}(6,q,k) = 6\times 3^{k-1}\\\midrule
        \addlinespace

        &\{7,q\} & \underline{\alpha} \to \alpha^r \underline{\beta} \alpha^t \beta \alpha^t \underline{\beta} \alpha^t \beta \alpha^t \underline{\beta} \alpha^r & \underline{\beta} \to \alpha^r \underline{\beta} \alpha^t \beta \underline{\alpha^t} \beta \alpha^t \underline{\beta} \alpha^r 
        &
        N_{\text{frac}}(7,q,k) = 7 \times 3^{k-1}
\\\midrule
        \addlinespace

        &\{8,q\} & \underline{\alpha} \to \alpha^r \underline{\beta} \alpha^t \beta  \alpha^t \underline{\beta} \underline{\alpha^t} \underline{\beta} \alpha^t \beta \alpha^t \underline{\beta} \alpha^r & \underline{\beta} \to \alpha^r \underline{\beta} \alpha^t \beta  \alpha^t \underline{\beta} \alpha^t \beta \alpha^t \underline{\beta} \alpha^r &
        N_{\text{frac}}(8,q,k) = 8 \times 4 ^{k-1}\\\midrule
        \addlinespace

        &\{9,q\} & \underline{\alpha} \to \alpha^r \underline{\beta} \alpha^t \beta  \alpha^t \underline{\beta} \alpha^t \beta  \alpha^t \underline{\beta} \alpha^t \beta \alpha^t \underline{\beta} \alpha^r & \underline{\beta} \to \alpha^r \underline{\beta} \alpha^t \beta  \alpha^t \underline{\beta} \underline{\alpha^t} \underline{\beta} \alpha^t \beta \alpha^t \underline{\beta} \alpha^r & 
        \text{No uniform pattern has been found} \\
        
        \bottomrule
    \end{tabular}
\end{table*}

\section{Conclusion}\label{sec:conc}

In this work, we have developed a comprehensive generalization of the hyperbolic fracton model (HFM) beyond the $\{5,4\}$ tessellation, establishing a unifying framework to understand fracton physics, holographic correspondence, and subsystem symmetry in negatively curved spaces. By systematically extending the model to generic $\{p,q\}$ tessellations, we revealed that the essential features of fracton physics, such as immobile excitations, \lu{extensive ground-state degeneracy, and  subsystem symmetries, persist in a wide class of hyperbolic geometries. The ground state degeneracy is finite and small for $q=3$, extensive for $q>3$ with the exception of \{4,4\} which is subextensive. In particular the ground state of the Euclidean \{6,3\} tessellation is also extensively degenerate. This suggests that the properties of the HFM are not only dictated by the negative curvature of the hyperbolic space, but also by the local connectivity $q$. In addition, for $p>4$ and $q>3$, the local structure of the tessellation enforces the number of fractons to grow exponentially in distance as they are expelled outwards, and algebraically in system size. Fractons of the HFM model are thus inherently different from type-I and type-II fractons on flat cubic lattices.}

The holographic nature of the model also manifests robustly across different tessellations. Through explicit Rindler reconstructions, we demonstrated that bulk information can be faithfully reconstructed from boundary subregions, satisfying subregion duality analogous to that in AdS/CFT. The Ryu-Takayanagi-like relation between mutual information and minimal surfaces was shown to hold quantitatively, confirming that the entanglement structure of the HFM encodes a discrete holographic duality. Finally, the introduction of a ``black hole'' region, implemented by excising a convex defect of ``deleted spins'', led to an entropy scaling proportional to the horizon perimeter, providing a natural realization of the black-hole area law in a purely lattice-based and locally constrained  system.

These results collectively establish that holography analogies in fracton models are not accidental but stem from a deeper geometric feature of  subsystem symmetries and the combinatorial structure of hyperbolic space. Our framework provides a controlled setting in which generalized symmetries, curvature, and \lu{mutual information} have intertwined relations. It opens the door to future investigations of quantum error correction, emergent gauge theories, and higher-rank tensor formulations on hyperbolic lattices. In particular, \lu{quantum dynamics may be added to our model in a way analogue to how the quantum X-cube model has been extended to $H^2\times S^1$ hyperbolic space with an extra spatial dimension of $S^1$ and proper boundary conditions.} Exploring the dynamical and finite-temperature behavior of the hyperbolic fracton phase, as well as its generalization to quantum fracton order, may illuminate new routes toward understanding the microscopic origin of holographic gravity in strongly correlated quantum matter, \lu{and} toward quantum error correction.

\hy{Finally, we comment on the relation between the hyperbolic lattice studied here and holographic constructions motivated by the AdS/CFT correspondence. The geometry considered in this work is the hyperbolic plane $H^2$, which may be viewed as a spatial time slice of AdS$_3$. However, it is important to emphasize that a simple vertical stacking of $H^2$ layers would instead produce the geometry $H^2\times\mathbb{R}$, which is not   equivalent to AdS$_3$. Our present construction therefore captures only the spatial structure of a hyperbolic slice, similar to many tensor-network realizations of holography defined on hyperbolic lattices. Incorporating a consistent dynamical time direction compatible with AdS geometry remains a nontrivial problem and lies beyond the scope of this work, but represents an interesting direction for future studies of fracton models in curved spacetimes.}

\section*{Acknowledgements} %The authors thank ... for useful discussions. 
Y.S. and L.J. acknowledge support from the French National Agency for Research (ANR-23-CE30-0038-01), Idex Bordeaux (Research Program GPR Light) and from the EUR Light S\&T (PIA3 Program, ANR-17-EURE-0027). H.Y. acknowledges the 2024 Toyota Riken Scholar Program from the Toyota Physical and Chemical Research Institute, the Overseas Research Support Grant from Yamada Science Foundation, and the  Grant-in-Aid for Research Activity Start-up from Japan Society for the Promotion of Science (Grant No. 24K22856).
 
\bibliography{biblio.bib}
\end{document}